\documentclass[journal]{IEEEtran}
\usepackage{hyperref}

\usepackage{cite}

\ifCLASSINFOpdf
\else
\fi
\usepackage{amsmath}
\usepackage{amssymb, bbm}
\ifCLASSOPTIONcompsoc
  \usepackage[caption=false,font=normalsize,labelfont=sf,textfont=sf]{subfig}
\else
  \usepackage[caption=false,font=footnotesize]{subfig}
\fi

\newtheorem{theorem}{Theorem}
\newtheorem{lemma}{Lemma}[section]
\newtheorem{defi}{Definition}[section]

\newtheorem{remark}{Remark}[section]

\newtheorem{example}{Example}[section]

\newcommand{\abs}[1]{\left \lvert #1\right \rvert}
\DeclareMathOperator*{\argmin}{arg\,min}
\DeclareMathOperator*{\argmax}{arg\,max}
\usepackage{mathtools}

\newcommand{\be}{\begin{equation}}
\newcommand{\ee}{\end{equation}}
\newcommand{\ben}{\begin{equation*}}
\newcommand{\een}{\end{equation*}}
\newcommand{\mc}{\mathcal}

\newcommand{\iid}{\text{i.i.d.}}
\newcommand{\stdnorm}{\mathcal{N}(0,1)}
\newcommand{\reals}{\mathbb{R}}
\newcommand{\expec}{\mathbb{E}}
\newcommand{\prob}{\mathbb{P}}

\newcommand{\SNR}{\textsf{snr}}
\newcommand{\mmse}{\textsf{mmse}}

\renewcommand\L{L}
\newcommand{\K}{K}
\newcommand{\M}{M}
\newcommand{\B}{B}
\newcommand{\Lc}{\textsf{C}}
\newcommand{\Lr}{\textsf{R}}
\newcommand{\sfr}{\textsf{r}}
\newcommand{\sfc}{\textsf{c}}

\newcommand{\pot}{\mc{F}}
\newcommand{\potargmin}{\mc{M}}

\newcommand{\bA}{\boldsymbol{A}}
\newcommand{\bS}{\boldsymbol{S}}
\newcommand{\bW}{\boldsymbol{W}}

\newcommand{\bXsec}{\boldsymbol{X}_\text{sec}}

\newcommand{\bZ}{\boldsymbol{Z}}

\newcommand{\bs}{\boldsymbol{s}}
\newcommand{\bw}{\boldsymbol{w}}
\newcommand{\bx}{\boldsymbol{x}}
\newcommand{\bz}{\boldsymbol{q}}
\newcommand{\by}{\boldsymbol{y}}
\newcommand{\bu}{\boldsymbol{u}}
\newcommand{\bv}{\boldsymbol{v}}

\newcommand{\bupsilon}{\boldsymbol{\upsilon}}
\newcommand{\btau}{\boldsymbol{\tau}}

\newcommand{\bpsi}{\boldsymbol{\psi}}

\newcommand{\indic}{ \mathbbm{1}}

\newcommand{\px}{p_{1}}
\newcommand{\e}{\epsilon}

\begin{document}
%
\title{Near-Optimal Coding for \\ Many-user Multiple Access Channels}
%
%
%

\author{Kuan~Hsieh,
        Cynthia~Rush,
        and~Ramji~Venkataramanan
\thanks{This work was supported in part by an Engineering and Physical Sciences Research Council Doctoral Training Partnership award. This paper was presented in part at the 2021 IEEE International Symposium on Information Theory.
K.~Hsieh and R.~Venkataramanan are with Department of Engineering, University of Cambridge, Cambridge CB2 1PZ, UK (e-mail: {\tt kh525@cam.ac.uk}; {\tt rv285@cam.ac.uk)}.
C.~Rush is with the Department of Statistics, New York, NY 10027, Columbia University, USA (e-mail: {\tt cynthia.rush@columbia.edu}).}
}

%
%


\maketitle

\begin{abstract}
This paper considers the Gaussian multiple-access channel in the asymptotic regime where the number of users grows linearly with the code length. We propose efficient coding schemes based on random linear models with approximate message passing (AMP) decoding and derive the asymptotic error rate achieved for a given user density, user payload (in bits), and user energy. The tradeoff between energy-per-bit and achievable user density (for a fixed user payload and target error rate) is studied.
It is demonstrated that in the large system limit, a spatially coupled coding scheme with AMP decoding achieves near-optimal tradeoffs for a wide range of user densities. Furthermore, in the regime where the user payload is large, we also study the tradeoff between energy-per-bit and  spectral efficiency and discuss methods to reduce decoding complexity.
\end{abstract}


%
\IEEEpeerreviewmaketitle

\section{Introduction}\label{sec:intro}
%
%
%
%
\IEEEPARstart{I}{n} certain modes of modern communications, such as in massive machine-type communications, a large number of devices simultaneously transmit to a single receiver. Furthermore, the data payload of each user (or device) in such applications may be small, e.g., temperature readings from a wireless sensor network.
These modes of communications do not 
lend themselves well to traditional large system analysis of multi-user systems, 
which often assume a fixed (small) number of users,
and the size of the user payload scaling linearly with the code length (as information rate is the metric of interest), see for example \cite[Chpt.~15]{cover2006elements} and \cite[Chpt.~4]{el2011network}.
This has motivated the study of multi-user channels in the \emph{many-user} or \emph{many-access} setting \cite{chen2017capacity}, where the number of users grows with the code length. This paper studies the asymptotic achievability of efficient coding schemes for the Gaussian multiple access channel (MAC) in the many-user setting.

The $\L$-user Gaussian MAC produces its 
output $\by \in \reals^{n}$ according to
\be
\label{eq:gmac}
\by = \sum_{\ell=1}^{L} \boldsymbol{c}_{\ell} + \bw,
\ee
where $\boldsymbol{c}_{\ell} \in \reals^n$ is the codeword of user $\ell \in \{1,\ldots, \L \}$ and the noise vector $\bw \in \reals^{n}$ has independent and identically distribution (i.i.d.) zero mean Gaussian entries with variance $\sigma^2>0$.

This paper first studies the symmetric Gaussian MAC in 
the asymptotic regime proposed by Polyanskiy et al.\ \cite{polyanskiy2017isit, zadik2019isit} where: 
(i) the number of users $\L$ grows linearly with the code length $n$, i.e., $\L = \mu n$ for some fixed user density $\mu\in(0,\infty)$,
(ii) the number of bits transmitted by each user (user payload), denoted by $\log_2 \M$,
is fixed and independent of $n$, and
(iii) the energy-per-bit, denoted by $E_b$, is fixed and independent of $n$. 
The energy needed to transmit each user's payload is $E=E_b \log_2 \M$.
The decoding metric used is the \emph{user error rate} (UER), which specifies the fraction of user messages decoded in error:
\be\label{eq:uer}
\text{UER} = \frac{1}{L} \sum_{\ell=1}^L \, \mathbbm{1} \left\{\hat{\bx}_{\ell} \neq \bx_{\ell}\right\},
\ee
where $\bx_{\ell}$ denotes the message sent by user $\ell$, 
and $\hat{\bx}_{\ell}$ is the decoder's estimate of the message. The \emph{per-user probability of error} (PUPE) error criterion used in \cite{polyanskiy2017isit, zadik2019isit} is the expected value of the UER.

In this asymptotic regime, the works of Polyanskiy et al.\ \cite{polyanskiy2017isit, zadik2019isit} obtained converse and achievability bounds on the minimum $E_b/N_0$ required to achieve a decoding error of $\text{PUPE}\leq \epsilon$ for a given $\epsilon>0$, when the user density $\mu$ and user payload $\log_2 \M$ are fixed.
Here $N_0=2\sigma^2$ is the noise spectral density. 
The achievability bound was based on the coding scheme where users encode their messages with i.i.d. Gaussian codebooks, and messages are decoded with (joint) maximum likelihood (ML) decoding.
We note that it is infeasible to implement the ML decoder in the many-user setting as its complexity scales exponentially with the number of users.

\subsection{Main Contributions and Structure of the Paper}
In this paper, we rigorously analyze the performance of coding schemes based on random linear models (which include i.i.d. Gaussian codebooks) and computationally efficient approximate message passing (AMP) decoding.
Our results provide the exact achievable regions of these schemes in the asymptotic regime described above, and demonstrate numerically that the achievable region of a coding scheme based on \emph{spatially coupled} Gaussian matrices and AMP decoding nearly matches the converse bound for a large range of user densities. An interesting feature of the spatially coupled scheme is that it can be interpreted as a block-wise time-division with overlap multiple-access scheme.

Section \ref{sec:random_linear_coding} introduces the random linear coding framework 
(based on either i.i.d.\ Gaussian or spatially coupled Gaussian matrices) 
and the associated AMP decoder. 
In  Section \ref{sec:asymptotic_uer}, we present Theorems \ref{thm:amp_perf} and \ref{thm:threshold_sat} which give the asymptotic UER achieved by these coding schemes for a fixed user payload. Then in Section \ref{sec:large_M} we bound the asymptotic UER achieved by these coding schemes as the user payload grows large (Theorems \ref{thm:amp_asympB} and \ref{thm:sc_amp_asympB}). 
Our results show that in the limit of large user payload, 
reliable communication is not possible at any fixed user density, 
and the interesting asymptotic regime is when the number of users $\L$, the code length $n$, and the user payload $\log_2 \M$ all tend to infinity with the spectral efficiency $S = \L \log_2 \M /n$ (total bits/channel use) held constant (Remark \ref{rmk:amp_asympB_spectraleff}). 
We also discuss how \emph{modulation} can be used in the proposed coding framework to reduce decoding complexity at large user payloads.

For simplicity, we only analyze coding schemes for the real-valued Gaussian MAC, but the results in this paper can be extended to the complex setting as well (see Section \ref{sec:large_M_implement}).


\subsection{Related Works}

The works of Chen et al.  \cite{chen2017capacity} and Ravi and Koch \cite{ravi2019isit, ravi2020IZS, ravi2021scaling} study MACs under different user scaling regimes. Moreover, their main results 
pertain to the setting where the probability of error goes to zero with increasing code length.
For the linear user scaling and finite error probability setting considered here, the recent papers \cite{kowshik2019isit, kowshik2021fundamental} study the fundamental tradeoffs in the quasi-static fading MAC. 
%

The AMP algorithms used for decoding in this paper (and the analyses of their asymptotic performance)
are similar to those used for random linear models and sparse superposition codes (SPARCs)\cite{donoho2009message, krzakala2012statistical, donoho2013information, joseph2012least, barbier2017approximate, barbier2016proof, rush2017capacity, rush2020capacity}.
The results in Section \ref{sec:asymptotic_uer} also generalize early results in random code-division multiple access (CDMA) \cite{tanaka2002cdma, guo2005randomly}. Indeed, the random linear coding framework described in Section \ref{sec:random_linear_coding} includes random CDMA as a special case.
%
%
%
The main novelty of our contribution is in applying AMP and spatial coupling techniques to 
design efficient coding schemes and prove that they achieve near-optimal tradeoffs for many-user MACs.  Our AMP analysis for many-user MAC differs from previous AMP analyses for random linear models and SPARCs in some key technical aspects. In our setting the components of the signal/message vector can be drawn from a probability distribution over finite length vectors, which is more general than in previous work. Furthermore, previous works use the potential function method to analyze the mean-squared error of AMP, whereas our results are in terms of the UER defined in \eqref{eq:uer}, which requires additional technical steps.
We note that  spatially coupled random CDMA and similar multiple-access schemes based on random spreading sequences have been studied under belief propagation decoding (with Gaussian approximations) in \cite{takeuchi2015performance,schlegel2013multiple}. However, these works do not analyze the many-user MAC setting considered in this paper.
 
M\"{u}ller \cite{muller2021gmac} recently proposed practical coding schemes for the many-user MAC setting considered here,
based on Gaussian random codes with power optimization and iterative soft decoding.
It was shown (non-rigorously) that the asymptotic spectral efficiency versus energy-per-bit tradeoff of the proposed scheme exceeds the achievability bound in \cite{zadik2019isit} at low spectral efficiencies, and is close to the converse bound in \cite{zadik2019isit} when the user payload is 100 bits.
As mentioned in \cite[Sec.~3.2]{muller2021gmac},   power optimized random codebooks with iterative decoding can be rigorously analyzed by generalizing the AMP analysis of power allocated SPARCs for (single-user) AWGN channels \cite{rush2017capacity}.
We leave this analysis and an in-depth  performance comparison of various many-user multiple-access schemes for future work.

Building on \cite{kowshik2021fundamental} and the first version of our paper, Kowshik \cite{kowshik2022improved} recently obtained a new achievability bound for many-user MACs based on spatially coupled random codebooks and a suboptimal AMP decoder. The advantage of this suboptimal decoder is that its analysis leads to a bound that can be easily computed for large user payloads (e.g., 100 bits). Our achievable region (computed using Theorem \ref{thm:threshold_sat}) is based on the optimal AMP denoiser, and though larger than the one in \cite{kowshik2022improved}, cannot be computed for  large payloads.

We emphasize that our problem setting is distinct from unsourced random access, 
where only a subset of users are active at any given time and all users use the same codebook (hence user messages are decoded only up to a permutation).
In our setting of the Gaussian MAC, users have different codebooks and are all active during the transmission period. Several recent works have analysed the fundamental tradeoffs and the performance of practical coding schemes for unsourced random access in the many-user setting \cite{polyanskiy2017isit, fengler2021sparcs, amalladinne2020approximate}.

\emph{Notation}: We use $\log_2$ and $\ln$ to denote the base 2 logarithm and natural logarithm, respectively.
The Gaussian distribution with mean $\mu$ and variance $\sigma^2$ is denoted by $\mc{N}(\mu, \sigma^2)$.
The indicator function of an event $\mc{A}$ is denoted by $\mathbbm{1}\{\mc{A}\}$.
For a positive integer $N$, we use $[N]$ to denote the set  $\{1, \dots, N\}$.
We use bold lower case letters or Greek symbols for vectors, and bold upper case for matrices and random vectors. 
We use plain font for scalars, and subscripts denote entries of a vector or matrix. For example, $\boldsymbol{x}$ denotes a vector, with $x_i$ being the $i^{th}$ element of $\boldsymbol{x}$. 
Similarly, the $(i,j)^{th}$ entry of matrix $\bA$ is denoted by $A_{ij}$.
The transpose of matrix $\bA$ is denoted by $\bA^*$. 
%

\section{Random linear coding and AMP decoding}\label{sec:random_linear_coding}

We consider coding schemes where the codewords of user $\ell \in [L]$ are constructed as $\boldsymbol{c}_\ell = \bA_\ell \bx_\ell$, where $\bA_\ell \in \reals^{n \times \B}$ is a random matrix and $\bx_\ell\in\reals^{\B}$ encodes the message of user $\ell$.
In this coding framework, the channel model \eqref{eq:gmac} can be written as
\be\label{eq:linear_regression}
\by = \bA\bx + \bw,
\ee
where the \emph{design matrix} $\bA\in\mathbb{R}^{n\times \L\B}$ is the horizontal concatenation of matrices $\bA_1, \ldots, \bA_\L$,
and the \emph{message vector} $\bx\in\mathbb{R}^{\L\B}$ is the concatenation of vectors $\bx_1, \ldots, \bx_\L$.
We will assume that the squared norm of each column of $\bA$ equals 1 in expectation.
For example, an i.i.d. Gaussian design matrix $\bA$ has i.i.d. $\mc{N}(0,\frac{1}{n})$ entries.

The $\L$ sections of $\bx$ (each of which corresponds to a user's message) are drawn i.i.d. from $p_{\bXsec}$, which is a probability mass function over a finite set of length $\B$ vectors.
 The per-user payload is therefore equal to the entropy $H(\bXsec)$, where $\bXsec \sim p_{\bXsec}$. The codeword energy constraint is denoted by $E = E_b \log_2 \M$, i.e., we require $p_{\bXsec}$ to satisfy $\expec\|\bXsec\|^2 = E < \infty$. 

\begin{example}[Random Codebooks]\label{ex:distr_unmod}
Let $p_{\bXsec}$ be the distribution over length $\B$ vectors that chooses uniformly at random one of its entries to be non-zero, taking the value $\sqrt{E}$. 
This corresponds to a user payload of $\log_2 M = \log_2 B$ bits.
User $\ell \in [\L]$ selects (and scales by $\sqrt{E}\,$) one-of-$\B$ columns from random matrix $\bA_\ell$, 
resulting in expected codeword energy $\expec\|\boldsymbol{c}_\ell\|^2 = E$.
In the rest of the paper, we denote the choice of $p_{\bXsec}$ used in this example by  $\px$.
\end{example}

\begin{example}[Random Codebooks with Binary Modulation]\label{ex:distr_binmod}
Let $p_{\bXsec}$ be the distribution over length $\B$ vectors that chooses uniformly at random one of its $\B$ entries to be non-zero, 
taking values in $\{\pm\sqrt{E}\}$ with equal probability. 
This corresponds to a user payload of $\log_2 \M =1 + \log_2 B$ bits.
User $\ell \in [\L]$ encodes $\log_2\B$ bits by selecting one-of-$\B$ columns from random matrix $\bA_\ell$, 
and an additional 1 bit in whether to modulate that column with $\sqrt{E}$ or $-\sqrt{E}$. When $\B=1$, this coding scheme corresponds to random CDMA with antipodal signalling.
\end{example}

One can generalize the distribution described in Example \ref{ex:distr_binmod} to consider other modulation schemes such as pulse amplitude modulation, or even complex modulation schemes such as phase-shift keying when the channel is complex (see Section \ref{sec:large_M_implement}).
Therefore, random linear coding with these choices of $p_{\bXsec}$ can be viewed as a generalization of random CDMA, with each user encoding $\log_2 \B$ bits in the choice of one-of-$\B$ random spreading sequences (in addition to the bits encoded in the choice of the modulation symbol).

\subsection{Spatially Coupled Coding Schemes}\label{sec:sc_coding}

\begin{figure*}[!t]
\centerline{
\subfloat[]{
	\includegraphics[width=0.55\textwidth]{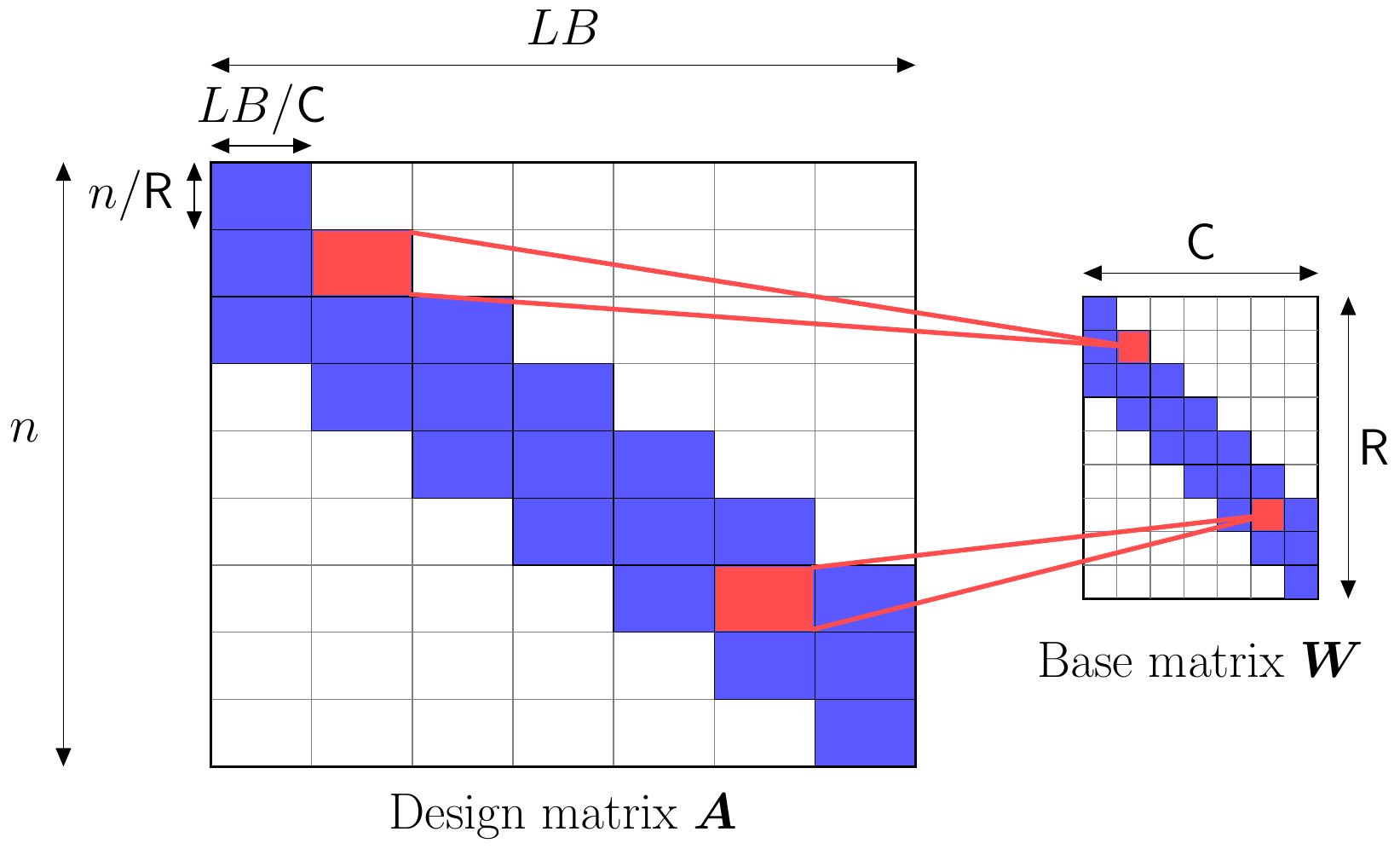}%
	\label{fig:sparc_scmatrix}}
\hspace{5em}
\subfloat[]{
	\includegraphics[width=0.28\textwidth]{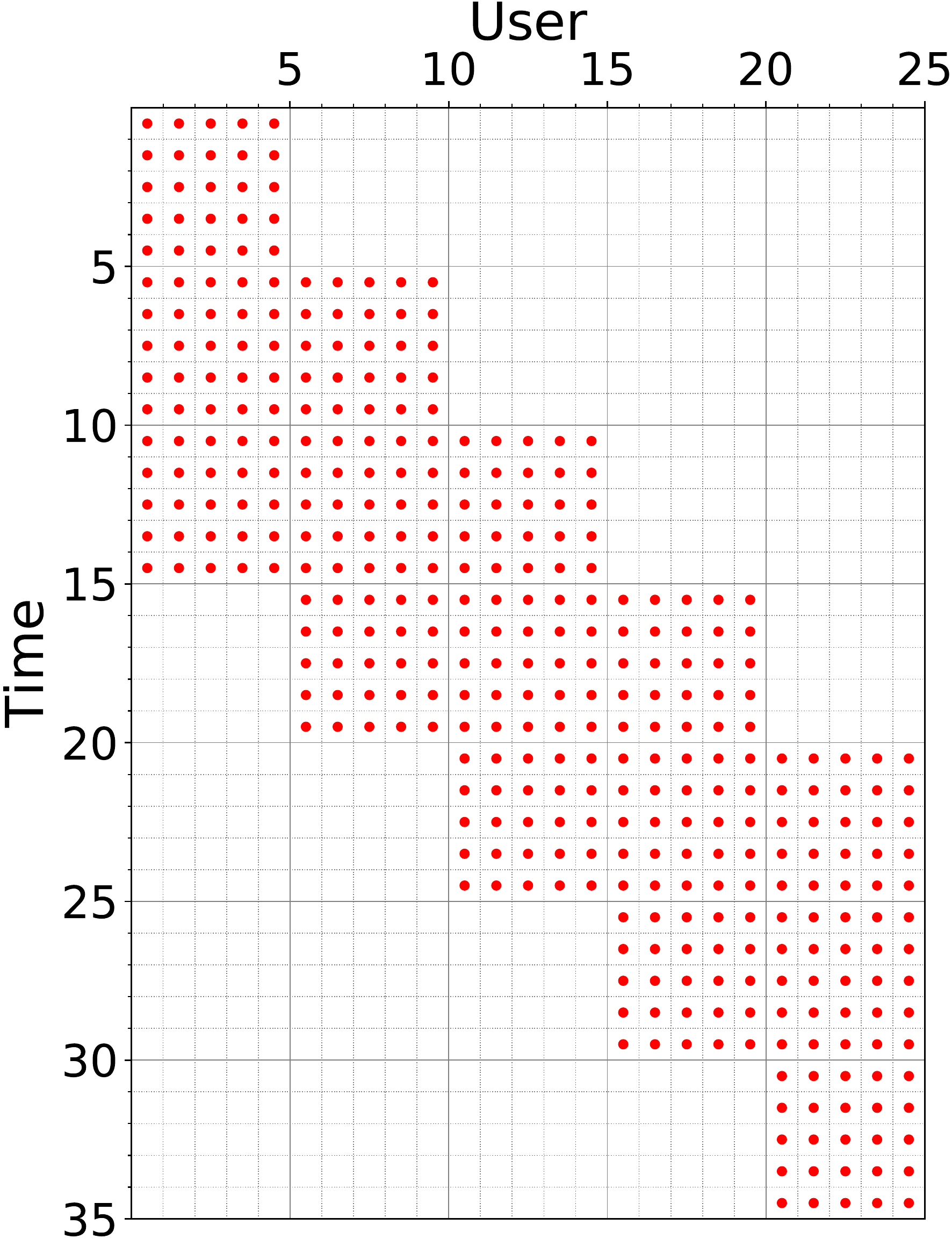}%
	\label{fig:time_user}}
}
\caption{(a) A spatially coupled design matrix $\bA$ defined using a base matrix $\bW \in \mathbb{R}_+^{9 \times 7}$. Each square in $\bW$ represents a scalar entry that specifies the variance of the entries in a block of $\bA$. The white parts of $\bA$ and $\bW$ correspond to zeros.
(b) An example of how 25 users communicate over 35 uses of the channel using the multiple access scheme based on a spatially coupled design matrix constructed using an $(\omega = 3, \Lambda = 5, \rho = 0)$ base matrix. A red dot in the 2D grid represents a certain user being active during a certain time instant and empty squares represent silence.
}
\label{fig:spatial_coupling}
\vspace{-5pt}
\end{figure*}

A spatially coupled (Gaussian) design matrix $\bA\in\mathbb{R}^{n\times\L\B}$ is divided into $\Lr$-by-$\Lc$ equally sized \emph{blocks}. The entries within each block are i.i.d. Gaussian with zero mean and variance specified by the corresponding entry of a base matrix $\bW \in \mathbb{R}_{+}^{\Lr \times \Lc}$.
The design matrix $\bA$ is constructed by replacing each entry of the base matrix $W_{\sfr \sfc}$ 
by an $\frac{n}{\Lr} \times \frac{\L\B}{\Lc}$ matrix with entries drawn i.i.d. from $\mc{N}(0, \frac{W_{\sfr \sfc}}{n/\Lr})$, 
for $\sfr\in[\Lr]$, $\sfc\in[\Lc]$.
See Fig. \ref{fig:sparc_scmatrix} for an example.
Hence,
the design matrix $\bA$ has independent Gaussian entries
\be\label{eq:construct_sc_A}
A_{ij} \sim \mc{N}\bigg(0,\frac{1}{n/\Lr} W_{\sfr(i) \sfc(j)} \bigg),  \quad \text{for }  i \in [n], \ j\in[\L \B].
\ee

The operators $\sfr(\cdot):[n]\rightarrow[\Lr]$ and $\sfc(\cdot):[\L\B]\rightarrow[\Lc]$ in \eqref{eq:construct_sc_A} map a particular row or column index to its corresponding \emph{row block} or \emph{column block} index. We require $\Lc$ to divide $\L$, resulting in $\L/\Lc\geq 1$ users per column block.
Recall that the design matrix $\bA\in\reals^{n \times \L\B}$ is the horizontal concatenation of the random matrices $\bA_1, \ldots, \bA_\L$ of all the users (see \eqref{eq:linear_regression}). 
Therefore, the $\ell$th section of the design matrix (columns $(\ell-1)\B+1$ to $\ell\B$) is the random matrix of user $\ell \in [\L]$.


The entries of the base matrix $\bW$ must satisfy 
$\sum_{\sfr=1}^{\Lr} W_{\sfr \sfc} = 1$ for $\sfc \in [\Lc]$
to ensure that the columns of the design matrix $\bA$ have unit norm in expectation.
The trivial base matrix with $\Lr=\Lc=1$ (single entry equal to 1) corresponds to the design matrix with i.i.d. $\mc{N}(0, \frac{1}{n})$ entries.
In this paper we will consider a class of base matrices called $(\omega, \Lambda, \rho)$ base matrices \cite{hsieh2018spatially, rush2020capacity}.

\begin{defi}
\label{def:ome_lamb_rho}
An $(\omega , \Lambda, \rho)$ base matrix $\bW$ is described by three parameters: coupling width $\omega\geq1$, coupling length $\Lambda\geq 2\omega-1$, and $\rho \in [0,1)$.
The matrix has $\Lr=\Lambda+\omega-1$ rows and $\Lc=\Lambda$ columns, with each column having $\omega$ identical non-zero entries in the band-diagonal. The  $(\sfr,\sfc)$th entry of the base matrix, for  $\sfr \in [\Lr], \sfc\in[\Lc]$, is given by
\be\label{eq:W_rc}
W_{\sfr \sfc} =
\begin{cases}
 	\ \frac{1-\rho}{\omega} \quad &\text{if} \ \sfc \leq \sfr \leq \sfc+\omega-1,\\
	\ \frac{\rho}{\Lambda-1} \quad &\text{otherwise}.
\end{cases}
\ee
\end{defi}

When $\rho=0$, as in Fig. \ref{fig:sparc_scmatrix},  the base matrix has non-zero entries only in the band-diagonal part. 
For example, the base matrix in Fig. \ref{fig:sparc_scmatrix} has parameters $(\omega=3, \Lambda=7, \rho=0)$.
For the simulations in Section \ref{sec:sc_gauss_sim}, we use $\rho = 0$, whereas for the theoretical results (Theorems \ref{thm:threshold_sat} and \ref{thm:sc_amp_asympB}) we choose $\rho$ to be a small positive value.
(The proofs of these results use the AMP concentration results in \cite{rush2020capacity}; there are some technical difficulties in proving the concentration results for $\rho=0$, which can be addressed by picking a suitable $\rho > 0$.)

Each entry of the base matrix corresponds to an $\frac{n}{\Lr} \times \frac{\L\B}{\Lc}$ block of the design matrix $\bA$, and each block can be viewed as an (uncoupled) i.i.d. Gaussian design matrix with $\frac{\L}{\Lc}$ users, code length $\frac{n}{\Lr}$, and user density
\be\label{eq:mu_inner}
\mu_\text{inner} = \frac{\L/\Lc}{n/\Lr} 
= \frac{\Lr}{\Lc} \, \mu 
= \Big(1 + \frac{\omega - 1}{\Lambda}\Big) \mu .
\ee
Since $\omega>1$ in spatially coupled systems, we have $\mu < \mu_\text{inner}$. This difference is often referred to as a ``rate loss'' in the literature of spatially coupled error correcting codes \cite{kudekar2011threshold, kudekar2013spatially, costello2014spatially, hsieh2018spatially}, and becomes negligible when $\Lambda$ is much larger than $\omega$.

The spatially coupled coding scheme can be viewed as \emph{block-wise time-division with overlap}.
Consider the scenario depicted in Fig. \ref{fig:time_user} where there are $\L=25$ users, $n=35$ channel uses, and a spatially coupled design matrix constructed using an $(\omega=3, \Lambda=5,\rho=0)$ base matrix $\bW\in \mathbb{R}_{+}^{7 \times 5}$.
Each block of the design matrix corresponds to 5 channel uses and 5 users.
Fig. \ref{fig:time_user} shows how users communicate using this multiple access scheme, assuming that each channel use corresponds to one time instant.
A red dot in the 2D grid represents a certain user being active (transmitting) during a certain time instant, and empty squares represent silence (no transmission). 
For example, users in the first column block (users 1 to 5) transmit during time instants 1 to 15 (corresponding to the first $\omega = 3$ row blocks) but are silent afterwards; 
users 6 to 10 transmit during time instants 6 to 20 but are silent otherwise, and so on.

Users within the same column block transmit simultaneously over $\omega$ row blocks of time (15 time instants), and users in neighbouring column blocks overlap in $\omega-1$ row blocks of time. 
At each time instant, up to $\omega$ column blocks of users (15 users) simultaneously transmit, but the set of active users gradually shifts over time. When $\omega= 1$, there is no time overlap (no coupling) between neighbouring blocks of users and each block of users communicates using an i.i.d. Gaussian matrix. 
When $\Lambda$ is large with respect to $\omega$, users are silent for most of the transmission period ($n$ channel uses). 
This facilitates low-complexity encoding and decoding.

With a band-diagonal spatially coupled matrix as in Fig. \ref{fig:sparc_scmatrix}, 
the sections at the two ends of the message vector $\bx$ are more easily decoded than other sections. This is because the channel outputs containing information about the sections at the ends have less interference from other sections (e.g., the first and last row of the base matrix $\bW$ only have one non-zero entry). Once the  sections at the ends have been decoded, their neighboring sections become easier to decode, and a decoding wave propagates towards the middle of the message vector. In contrast, for i.i.d. design matrices, all sections are equally hard to decode. 
For more intuitive explanations and figures that illustrate the decoding wave in spatially coupled systems, see \cite{krzakala2012statistical, donoho2013information, barbier2017approximate, rush2020capacity}.


\subsection{AMP Decoding and State Evolution}\label{sec:scamp}

We consider an efficient AMP decoder that aims to reconstruct the message vector $\bx$ from the channel output $\by$. 
The decoder knows the design matrix $\bA$, 
the base matrix $\bW$,
the distribution $p_{\bXsec}$, and the 
channel noise variance $\sigma^2$.
AMP algorithms are based on Gaussian approximations to loopy belief propagation on dense graphs \cite{donoho2009message} 
and have been proposed for estimation in the random linear model \eqref{eq:linear_regression} with spatially coupled design matrices $\bA$ \cite{krzakala2012statistical, donoho2013information, barbier2017approximate, rush2020capacity}.

The AMP decoder iteratively produces message vector estimates $\bx^{t}\in\mathbb{R}^{\L\B}$ for iterations $t=1,2,\ldots$ as follows.
Initialize $\bx^0$ to the all-zero vector, and for $t \geq 0$, iteratively compute:
\begin{align}\label{eq:sc_amp}
\begin{split}
\bz^t &= \by - \bA \bx^t + \widetilde{\bupsilon}^t \odot \bz^{t-1},\\
\bx^{t+1} &= \eta^t \big(\bx^t + (\widetilde{\bS}^t \odot \bA)^*  \bz^t\big).
\end{split}
\end{align}
Here $\odot$ is the Hadamard (entry-wise) product and quantities with negative iteration indices are set to zero vectors.
The vector $\widetilde{\bupsilon}^t\in\reals^{n}$, 
the matrix $\widetilde{\bS}\in\reals^{n\times\L\B}$,
and the denoising function $\eta^t$ will be described in terms of the following state evolution parameters.

\textit{State Evolution:}
The performance of the AMP in the large system limit  is succinctly captured by a deterministic recursion called \emph{state evolution} \cite{bayati2011thedynamics, javanmard2013state}.
State evolution iteratively defines vectors $\boldsymbol{\gamma}^t, \boldsymbol{\phi}^t\in\mathbb{R}^{\Lr}$ and $\boldsymbol{\tau}^t, \boldsymbol{\psi}^t\in\mathbb{R}^{\Lc}$ as follows. 
Initialize $\psi_\sfc^0 = E$ for $\sfc \in [\Lc]$, and for $t\geq 0$, iteratively compute:
\begin{align}
	\gamma_\sfr^t &= \sum_{\sfc=1}^{\Lc} W_{\sfr \sfc} \psi_\sfc^t , \hspace{0.9cm}
	\phi_\sfr^t = \sigma^2 + \mu_{\text{inner}}  \gamma_\sfr^t , \hspace{0.78cm} \sfr\in[\Lr], \label{eq:sc_se_phi}\\
	\tau_\sfc^t &= \bigg[\sum_{\sfr=1}^{\Lr}\frac{W_{\sfr \sfc}}{\phi_\sfr^t}\bigg]^{-1}, \hspace{1em}
	\psi_\sfc^{t+1} = \mmse(1/\tau_\sfc^t), \hspace{1em} \sfc\in [\Lc], \label{eq:sc_se_psi}
\end{align}
where $\mu_{\text{inner}} = \frac{\Lr}{\Lc} \, \mu $ from \eqref{eq:mu_inner}, 
and
\begin{align}
\mmse(1/\tau)
&= \expec \, \Big\|\bXsec - \expec\Big[\bXsec  \mid \bXsec + \sqrt{\tau}\bZ\Big] \Big\|^2\label{eq:mmse_func}\\
&\stackrel{\text{(i)}}{=} E \, \Bigg[1 - \mathbb{E}\Bigg[ \frac{e^{\sqrt{\frac{E}{\tau}} Z_1}}{e^{\sqrt{\frac{E}{\tau}} Z_1} + e^{ {-E}/{\tau}}\sum_{j=2}^{\B} e^{\sqrt{\frac{E}{\tau}} Z_j} }\Bigg] \Bigg],\nonumber
\end{align}
where $\bXsec \sim p_{\bXsec}$ and $\bZ = [Z_1, \ldots, Z_\B]$ is a standard Gaussian vector independent of $\bXsec$.
The equality (i) holds when $\bXsec\sim \px$, where $\px$ is described in Example \ref{ex:distr_unmod}.

The vector $\widetilde{\bupsilon}^t \in \mathbb{R}^n$ and the matrix $\widetilde{\bS}^t \in \mathbb{R}^{n\times \L\B}$ in \eqref{eq:sc_amp} both have a block-wise structure and are defined using state evolution parameters as follows.
For $i\in[n]$ and $j\in[\L\B]$,
\begin{align}\label{eq:sc_amp_se_params}
	\widetilde{\upsilon}_i^t = \frac{\mu_{\text{inner}} \,\gamma_{\sfr(i)}^t}{\phi_{\sfr(i)}^{t-1}},
	\qquad \widetilde{S}_{ij}^t = \frac{\tau_{\sfc(j)}^t}{\phi_{\sfr(i)}^t},
\end{align}
where we recall that $\sfr(i)$ and $\sfc(j)$ denote the row and column block index of the $i$th row entry and $j$th column entry, respectively.
The vector $\widetilde{\bupsilon}^0$ is defined to be the all-zero vector.

In each iteration, the AMP decoder \eqref{eq:sc_amp} produces an effective observation 
$\bs^t = \bx^t + (\widetilde{\bS}^t \odot \bA)^*  \bz^t$,
which has the following approximate representation: for an index $j$ in column block $\sfc$ of the message vector $\bx$, we have $s^t_j \approx x_j +  \sqrt{\tau^t_{\sfc}} Z_j$, where $\{ Z_j  \}\sim_{\text{i.i.d.}} \mc{N}(0,1)$. 
The estimate $\bx^{t+1}$ in \eqref{eq:sc_amp} is then the minimum mean square error (MMSE) estimate of $\bx$ given $\bs^t$, computed using the assumed distribution. 
This leads to the following definition of 
the denoising function $\eta^t = (\eta^t_1, \ldots, \eta^t_{\L\B})$ in \eqref{eq:sc_amp}:
for index $j$ in section $\ell \in [\L]$, which we denote by $j\in\sec(\ell)$,
\begin{align}
\eta^t_j(\bs)
&= \mathbb{E}\left[(\bXsec)_j \mid \bXsec + \sqrt{\tau^t_{\sfc(j)}}\, \bZ = \bs_\ell \right]
\label{eq:eta_def}\\
&\stackrel{\text{(i)}}{=} \sqrt{E} \cdot \frac{\exp\Big({s_j  \sqrt{E}}/{\tau^t_{\sfc(j)}}\Big)}{\sum_{i \in \sec(\ell)}
\exp\Big({s_{i}  \sqrt{E}}/{\tau^t_{\sfc(j)}}\Big)},\label{eq:eta_def_p1}
\end{align}
where we recall that $\bs_\ell \in \reals^{\B}$ is the $\ell$th section of  $\bs \in\reals^{\L\B}$.
The equality (i) holds when $\bXsec\sim \px$.

In addition, the decoder can also produce a \emph{hard-decision} maximum a posteriori (MAP) estimate from $\bs^t$, which we denote by $\hat{\bx}^{t+1}$. 
For section $\ell$ in column block $\sfc\in[\Lc]$, 
the $\ell$th section of this hard-decision estimate is given by
\be\label{eq:MAPdef}
\hat{\bx}^{t+1}_{\ell}
= \argmax_{\bx' \in \mc{X}} \, \mathbb{P}\Big( \bXsec = \bx' \mid \bXsec + \sqrt{\tau^t_{\sfc}}\,\bZ = \bs^t_\ell\Big),
\ee
where $\mc{X}$ is the support of $p_{\bXsec}$.
When $\bXsec\sim \px$, for $j \in \sec(\ell)$ this hard-decision estimate is given by
\begin{align}
\hat{x}^{t+1}_j
&=\begin{cases}
 	\ \sqrt{E} \quad &\text{if} \ s^t_j > s^t_{i}  \text{ for all }  i \in \sec(\ell)\backslash j,\\
	\ 0 \quad &\text{otherwise}.
\end{cases}\label{eq:amp_final_hard}
\end{align}

Eqs. \eqref{eq:eta_def_p1} and \eqref{eq:amp_final_hard} 
give closed form expressions for the denoising function $\eta^t$ and MAP estimator when $\bXsec\sim \px$.
Similar expressions can be easily obtained when $\bXsec$ is drawn from the discrete distribution in Example \ref{ex:distr_binmod} or its generalizations, but these are omitted for brevity.
%

\textit{IID Gaussian $\bA$:} For the special case where the entries of the design matrix $\bA$ are  i.i.d. $\sim \mc{N}(0,\frac{1}{n})$, the AMP decoder \eqref{eq:sc_amp} and the state evolution \eqref{eq:sc_se_phi}--\eqref{eq:sc_se_psi} can be simplified.
Setting $\bx^0$ equal to the all-zero vector, for $t \geq 0$,  the AMP decoder computes:
\begin{align}\label{eq:amp}
\begin{split}
\bz^t &= \by - \bA \bx^t + \frac{\mu \psi^t}{\tau^{t-1}} \, \bz^{t-1},\\
\bx^{t+1} &= \eta^t \big(\bx^t + \bA^*  \bz^t\big).
\end{split}
\end{align}
At $t=0$, the vector $\frac{\mu \psi^0}{\tau^{-1}}  \bz^{-1}$ is set to the all-zero vector.
%
The scalars $\tau^t$ and $\psi^t$ are computed via the state evolution recursion.
Initializing with $\psi^0 = E$, for $t \geq 0$ we have:
\be
\begin{split}
\label{eq:se}
	\tau^t &= \sigma^2 + \mu  \psi^t, \\
	\psi^{t+1} &= \mmse(1/\tau^t), 
\end{split}
\ee
where the $\mmse$ function is defined in \eqref{eq:mmse_func}.
Furthermore, in this case,
the denoising function $\eta^t$ in \eqref{eq:eta_def}
and the hard-decision estimate $\hat{\bx}^{t+1}$ in \eqref{eq:MAPdef} 
are defined using the state evolution parameter $\tau^t$ as the Gaussian noise variance.

\section{Asymptotic UER achieved by AMP decoding}\label{sec:asymptotic_uer}

We now characterize the asymptotic UER (see \eqref{eq:uer}) achieved by coding schemes based on i.i.d. and spatially coupled Gaussian design matrices with AMP decoding.
These results are stated in terms of a \emph{potential function}.

\subsection{Potential Function}\label{sec:pot_func}
Consider the single-section Gaussian channel with noise variance $\tau$:
\be\label{eq:single_user_ch}
\bS_{\tau} = \bXsec + \sqrt{\tau} \bZ,
\ee
where $\bXsec \sim p_{\bXsec}$ and $\bZ\in \reals^{\B}$ is a standard Gaussian vector independent of $\bXsec$.
The potential function for the random linear system \eqref{eq:linear_regression} with user density $\mu = \frac{\L}{n}$ and channel noise variance $\sigma^2$ is defined as
\begin{align}\label{eq:potential_func}
\pot(\mu,\sigma^2, \psi) 
= I(\bXsec ;  \bS_{\tau}) 
+ \frac{1}{2\mu} \bigg[\ln\Big(\frac{\tau}{\sigma^2}\Big) - \frac{\mu\psi}{\tau}\bigg],
\end{align}
where $\psi\in[0,E]$, $\tau = \sigma^2 + \mu \psi$, and the mutual information $I(\bXsec ;  \bS_{\tau})$ is computed using the channel \eqref{eq:single_user_ch}.%
\footnote{The potential function \eqref{eq:potential_func} has connections with the mutual information between the message vector and the channel output vector in the random linear estimation problem \eqref{eq:linear_regression} \cite{reeves2019thereplica, barbier2020mutual}. 
We do not use this connection here and only consider the relationship between the stationary points of the potential function and the fixed points of the state evolution recursion. The potential functions used
in this paper can be derived using the method in \cite{yedla2014asimple}, up to scaling factors and additive  constants (which do not affect the desired properties).
}
If $\bXsec \sim \px$ and $\bZ = [Z_1, \ldots, Z_\B]$, then
\begin{align*}
I (\bXsec ;  \bS_{\tau}) 
= \frac{E}{\tau} + \ln B 
- \expec \ln \bigg[ e^{\frac{E}{\tau}  + \sqrt{\frac{E}{\tau}} Z_1}+ \sum_{j=2}^B  e^{\sqrt{\frac{E}{\tau}} Z_j} \bigg].
\end{align*}

Define the set of potential function minimizers (w.r.t. $\psi$) as:
\be\label{eq:potential_func_argmin}
\potargmin(\mu,\sigma^2)
= \argmin_{\psi \in [0, E]} \, \pot(\mu,\sigma^2, \psi).
\ee

Consider decoding $\bXsec$ from $\bS_{\tau}$ produced by the Gaussian channel in \eqref{eq:single_user_ch}.
The MMSE decoder
$\hat{\bx}_{\text{sec}}^{\text{MMSE}}(\bS_{\tau}) = \expec[\bXsec | \bS_{\tau}]$
achieves the MMSE given by \eqref{eq:mmse_func}.
The MAP decoder 
$\hat{\bx}_{\text{sec}}^{\text{MAP}}(\bS_{\tau}) = \argmax_{\bx'} \prob(\bXsec = \bx'| \bS_{\tau})$
achieves the minimum probability of error given by
\begin{align}
P_e(\tau) 
&= \mathbb{P} \big(\hat{\bx}_{\text{sec}}^{\text{MAP}}(\bS_{\tau}) \neq \bXsec\big) \label{eq:Pe}\\
&\stackrel{\text{(i)}}{=} 1 - \mathbb{E} \bigg[\Phi\Big(\sqrt{E/\tau} + Z\Big)^{B-1} \bigg].\label{eq:Pe_unmod}
\end{align}
where $Z\sim\stdnorm$ and $\Phi(\cdot)$ is the standard normal distribution function. The equality (i) holds when $\bXsec\sim \px$.
%

Theorem 1 below shows that for i.i.d. Gaussian matrices, the fixed point of the state evolution equations \eqref{eq:se} is characterized by a specific stationary point of the potential function \eqref{eq:potential_func},
and that the asymptotic UER achieved by AMP decoding can be  bounded using the fixed point. 
Theorem 2 gives an analogous result for spatially coupled Gaussian matrices, where the fixed point of the (spatially coupled) state evolution \eqref{eq:sc_se_phi}--\eqref{eq:sc_se_psi} is bounded using the global minimum of the potential function.

\subsection{IID Gaussian Matrices}

\begin{theorem}[IID Gaussian Matrices with AMP Decoding]\label{thm:amp_perf}
Consider the linear model \eqref{eq:linear_regression}, with the entries of the design matrix $\bA$  i.i.d. $\sim \mc{N}(0,\frac{1}{n})$ and the $L$ sections of the message vector $\bx$  i.i.d. $\sim p_{\bXsec}$.
Let $\hat{\bx}^t$ be the AMP hard-decision estimate of $\bx$ after iteration $t$ (see \eqref{eq:MAPdef}), 
and recall that $\tau^t, \psi^t$ are outputs of the state evolution \eqref{eq:se}.

1) The sequences $\{\tau^t\}_{t\geq0}$ and $\{\psi^t\}_{t\geq0}$ are non-increasing and converge to fixed points $\tau^{\text{FP}}$, $\psi^{\text{FP}}$, where 
\begin{align}
\tau^{\text{FP}} &\coloneqq \sigma^2 + \mu \psi^{\text{FP}}, \label{eq:se_fp_pot_tau}\\
\psi^{\text{FP}} 
&\coloneqq \max \bigg\{\psi : \psi=\mmse\bigg(\frac{1}{\sigma^2 +\mu\psi}\bigg)\bigg\}\nonumber\\
&=\max \bigg\{\psi : \frac{\partial \pot(\mu,\sigma^2,\psi)}{\partial \psi}=0\bigg\}. \label{eq:se_fp_pot_psi}
\end{align}
The potential function $\pot(\mu,\sigma^2,\psi)$ is defined in \eqref{eq:potential_func}.

2) Fix $\delta>0$, and let $T$ denote the first iteration for which $\tau^t \leq \tau^{\text{FP}} + \delta$.
Then the UER of the AMP decoder after $T+1$ iterations satisfies the following almost surely:
\be\label{eq:amp_se_convergence}
\lim_{\L\to\infty} 
\frac{1}{\L} \sum_{\ell=1}^\L \mathbbm{1} \big\{\hat{\bx}_\ell^{T+1} \neq \bx_\ell \big\} 
\stackrel{\text{a.s.}}{=} P_e(\tau^T)
\leq P_e(\tau^{\text{FP}} + \delta),
\ee
where the limit is taken with $\frac{\L}{n}=\mu$ held constant and $P_e(\cdot)$ is defined in \eqref{eq:Pe}.
\end{theorem}

\begin{IEEEproof}
1) Results similar to the first part of the theorem are known in the AMP literature and are sometimes used implicitly. 
We provide a proof here for completeness. We first prove that the sequence $\{\psi^t\}_{t\geq0}$ is non-increasing and converges to the fixed point $\psi^{\text{FP}}$ defined in \eqref{eq:se_fp_pot_psi}.
Then the result for $\{\tau^t\}_{t\geq0}$ immediately follows since $\tau^t = \sigma^2 + \mu\psi^t$.

Writing the state evolution in \eqref{eq:se} as a single recursion, we have: 
\be\label{eq:se_single}
\psi^{t+1}=\mmse((\sigma^2 +\mu\psi^t)^{-1}). 
\ee
Starting from $\psi^0=\expec\|\bXsec\|^2=E$, we have that 
\be\label{eq:psi1_leq_psi0}
\psi^1 = \mmse((\sigma^2 +\mu E)^{-1}) \leq E = \psi^0,
\ee
where the inequality holds because the trivial all zero estimate of a random section $\bXsec$ achieves an expected squared error of $E$.
The $\mmse$ function defined in \eqref{eq:mmse_func} is non-increasing  in its argument \cite{guo2011estimation, barbier2016proof}, and since its argument $(\sigma^2 +\mu\psi^t)^{-1}$ is decreasing in $\psi^t$, the $\mmse$ function is non-decreasing in $\psi^t$. 
Therefore, 
together with \eqref{eq:psi1_leq_psi0}, this shows that the sequence $\{\psi^t\}_{t\geq0}$ is non-increasing.
%
%
Moreover, if $\psi^t \geq \psi^{\text{FP}}$, then $\psi^{t+1}\ge  \psi^{\text{FP}}$. 
Indeed, for any $\psi^t \geq \psi^{\text{FP}}$,
\begin{equation*}
\begin{split}
\psi^{t+1} 
&= \mmse((\sigma^2 +\mu\psi^t)^{-1})\\
&\geq \mmse((\sigma^2 +\mu\psi^{\text{FP}})^{-1})
=\psi^{\text{FP}}.
\end{split}
\end{equation*}
Since $\{\psi^t\}_{t\geq0}$ is a non-increasing sequence bounded below by $\psi^{\text{FP}}$ (noting that $\psi^0\geq\psi^{\text{FP}}$), we conclude that it converges to $\psi^{\text{FP}}$.
%

To show that the fixed points of the state evolution correspond to the stationary points of the potential function $\pot(\mu,\sigma^2,\psi)$ defined in \eqref{eq:potential_func}, we compute the derivative:
\begin{align}
\frac{\partial \pot(\mu,\sigma^2,\psi)}{\partial \psi}
&= \frac{\mu}{2(\sigma^2 + \mu\psi)^2}  \, \bigg[\psi - \mmse\bigg(\frac{1}{\sigma^2+\mu\psi}\bigg)\bigg],\nonumber
\end{align}
where we have used $\tau = \sigma^2 + \mu \psi$
and the vector I-MMSE relationship \cite[Thm.~2]{guo2005mutual}.
Therefore, since $\sigma^2>0$ and $\mu>0$, we have that $\partial \pot(\mu,\sigma^2,\psi)/\partial \psi=0$ corresponds to $\psi = \mmse ((\sigma^2+\mu\psi)^{-1})$, 
which is the fixed point of the iteration \eqref{eq:se_single}.

2) We now prove \eqref{eq:amp_se_convergence}. 
For $\ell \in [L]$, we denote by $\boldsymbol{a}_\ell \in \reals^B$  the $\ell$th section of a vector $\boldsymbol{a} \in \reals^{LB}$.   
Consider the input to the AMP hard-decision step in iteration $t+1$,
which we denote by $\bs^t = \bx^t +\bA^* \bz^t \in\reals^{\L\B}$ (see \eqref{eq:MAPdef}, \eqref{eq:amp}).
The MAP estimator $\hat{\bx}^{t+1}_\ell = \hat{\bx}^{t+1}_\ell(\bs^t_\ell)$ in \eqref{eq:MAPdef} partitions the space $\reals^B$ into decision regions. For each $\bx_\ell$ in the support of $p_{\bXsec}$, the decision region is
\be\label{eq:MAPdecregion}
\mc{D}(\bx_\ell) := \left\{ \bs^t_\ell : \, \hat{\bx}^{t+1}_\ell(\bs^t_\ell) = \bx_\ell  \right\}.
\ee
Note that $\indic\{ \hat{\bx}^{t+1}_\ell(\bs^t_\ell)= \bx_\ell \} = \indic\{ \bs^t_\ell \in \mc{D}(\bx_\ell) \}$.

The distance between a vector $\bv \in \reals^B$  and a set $\mc{B} \subset \reals^{B}$ is denoted by 
$d(\bv, \mc{B}) := \inf\{ \| \bv - \bu \|_2: \, \bu \in \mc{B} \}$. For any $\e >0$, define the functions 
$\xi_{\e,+}, \xi_{\e,-}: \ \reals^{B} \times \reals^{B} \to \reals$ as follows:
\begin{align*} 
\xi_{\e,+}(\bx_\ell, \bs^t_\ell) =
\begin{cases}
1, & \text{ if } \, \bs^t_\ell \in \mc{D}(\bx_\ell), \\
0, & \text{ if } \, d(\bs^t_\ell, \mc{D}(\bx_\ell))>\e, \\
1-  d(\bs^t_\ell, \mc{D}(\bx_\ell))/\e, & \text{ otherwise},
\end{cases}
\end{align*}
\begin{align*} 
\xi_{\e,-}(\bx_\ell, \bs^t_\ell) =
\begin{cases}
1, & \text{ if } \, d(\bs^t_\ell, \mc{D}(\bx_\ell)^c) > \e, \\
0, & \text{ if } \, \bs^t_\ell \in \mc{D}(\bx_\ell)^c, \\
 d(\bs^t_\ell, \mc{D}(\bx_\ell)^c)/\e, & \text{ otherwise}.
\end{cases}
\end{align*}
We note that $\xi_{\e,+}, \xi_{\e,-}$ are Lipschitz-continuous (with Lipschitz constant  $1/\e$), and
\[ \xi_{\e,-}(\bx_\ell, \bs^t_\ell)  \leq  \indic\{ \bs^t_\ell \in \mc{D}(\bx_\ell) \} \leq \xi_{\e,+}(\bx_\ell, \bs^t_\ell),  \]
and thus
\be
\label{eq:perr_ub_lb}
\begin{split}
\frac{1}{L}\sum_{\ell=1}^L \xi_{\e,-}(\bx_\ell, \bs^t_\ell)  & \leq 
\frac{1}{L}\sum_{\ell=1}^L
\indic\{ \hat{\bx}^{t+1}_\ell(\bs^t_\ell) = \bx_\ell \}  \\
& \leq   \frac{1}{L}\sum_{\ell=1}^L \xi_{\e,+}(\bx_\ell, \bs^t_\ell).
\end{split}
\ee

A pseudo-Lipschitz function $\xi: \reals^{m}  \to \reals$ is one that satisfies the following for all $\bu, \bv \in \reals^{m}$:
\[ \abs{\xi(\bu) - \xi(\bv)} \le C(1 + 
\| \bu \|_2 + \| \bv \|_2)\, 
\| \bu - \bv\|_2,   \]
for some constant $C >0$.
The results in \cite{bayati2011thedynamics} and  \cite{rush2017capacity,rush2019theerror,rush2020capacity} imply that for any pseudo-Lipschitz function $\xi: \reals^{B} \times \reals^{B} \to \reals$, the following holds almost surely:
\be \label{eq:PL2_res}
\lim_{L \to \infty} \frac{1}{L} \sum_{\ell=1}^{L} \xi(\bx_\ell, \bs^t_\ell ) = \expec\{ \xi( \bXsec, \, \bS_{\tau^t} ) \},
\ee
where $\bXsec \sim p_{\bXsec}$ and $\bS_{\tau^t}$ is given by \eqref{eq:single_user_ch}. 
This result was proved in \cite{bayati2011thedynamics} for the $B=1$ case and extended in \cite{rush2017capacity,rush2020capacity}  to the setting of SPARCs where the specific distribution $p_{\bXsec}$ given in Example \ref{ex:distr_unmod} (corresponding to random codebooks) is used. The proof for more general discrete distributions is essentially the same. 
In \eqref{eq:PL2_res} and  below, $\L/n= \mu$ as $\L \to \infty$.

Applying \eqref{eq:PL2_res} to the Lipschitz continuous functions $\xi_{\e,+}$ and $\xi_{\e,-}$, we obtain almost surely:
\be
\begin{split}
\label{eq:epm_limits}
& \lim_{L \to \infty } \frac{1}{L} \sum_{\ell=1}^{L} \xi_{\e, +}(\bx_\ell, \bs^t_\ell ) = \expec\{ \xi_{\e,+}( \bXsec, \, \bS_{\tau^t} ) \},  \\
& \lim_{L \to \infty } \frac{1}{L} \sum_{\ell=1}^{L} \xi_{\e, -}(\bx_\ell, \bs^t_\ell ) = \expec\{ \xi_{\e,-}( \bXsec, \, \bS_{\tau^t} ) \}.
\end{split}
\ee
Since $\e > 0$ is arbitrary, from \eqref{eq:perr_ub_lb} and \eqref{eq:epm_limits}, we almost surely have
\begin{equation*}
\begin{split}
 \lim_{\e \to 0 } \expec\{ \xi_{\e,-}( \bXsec, \, \bS_{\tau^t} ) \}  
&\leq  \liminf_{L \to \infty} \frac{1}{L}\sum_{\ell=1}^L
\indic\{ \hat{\bx}^{t+1}_\ell(\bs^t_\ell) = \bx_\ell \} \\
& \leq  \limsup_{L \to \infty} \frac{1}{L}\sum_{\ell=1}^L
\indic\{ \hat{\bx}^{t+1}_\ell(\bs^t_\ell) = \bx_\ell \}  \\
&  \leq \lim_{\e \to 0 } \expec\{ \xi_{\e,+}( \bXsec, \, \bS_{\tau^t} ) \}.
\end{split}
\end{equation*}
By the dominated convergence theorem, we have
\begin{equation*}
\begin{split}
& \lim_{\e \to 0 } \expec\{ \xi_{\e,-}( \bXsec, \, \bS_{\tau^t} ) \}  = \prob( \bS_{\tau^t} \in \mc{D}(\bXsec) )= 1 -P_e(\tau^t), \\ 
& \lim_{\e \to 0 } \expec\{ \xi_{\e,+}( \bXsec, \, \bS_{\tau^t} ) \}  = \prob(\bS_{\tau^t} \in \mc{D}(\bXsec) )= 1 -P_e(\tau^t).
\end{split}
\end{equation*}
This completes the proof that, almost surely
\be \label{eq:MAP_Pe_conv}
 \lim_{L\to \infty} \frac{1}{L}\sum_{\ell=1}^L
\indic\{ \hat{\bx}^{t+1}_\ell(\bs^t_\ell) \neq \bx_\ell \} = P_e(\tau^t).
\ee
\end{IEEEproof}

\begin{remark}\label{rem:replica}
Consider the setting of Theorem \ref{thm:amp_perf} and the section-by-section (SBS) MAP decoder for the linear model  
\eqref{eq:linear_regression}: 
\be
\hat{\bx}^{\text{MAP}}_\ell
= \argmax_{\bx' \in \mc{X}} \prob(\bx_\ell = \bx' \mid \by, \bA), \quad \text{for } \ell \in [L],
\label{eq:SBS_MAP}
\ee
where $\mc{X}$ is the support of the discrete prior for each section of the message vector.
The SBS-MAP decoder in \eqref{eq:SBS_MAP} minimizes the expected UER.
%
Though computationally infeasible, the asymptotic error of the SBS-MAP decoder can be analyzed using the non-rigorous replica method.
Using this technique, Tanaka \cite{tanaka2002cdma} showed that for binary CDMA ($B=1$ and $p_{\bXsec}$ uniform over $\{1, -1 \}$), the asymptotic UER of $\hat{\bx}^{\text{MAP}}$
can be characterized in terms of the probability of decoding error in the single-section Gaussian channel \eqref{eq:single_user_ch}.
Specifically, when $\potargmin(\mu,\sigma^2)$ in \eqref{eq:potential_func_argmin} is a singleton (i.e., when the global minimizer of the potential function is unique),
\be
\lim_{\L\to\infty} \frac{1}{\L} \sum_{\ell=1}^\L \indic\{  \hat{\bx}_\ell^{\text{MAP}} \neq \bx_\ell \} \label{eq:replica_pupe}
= P_e(\tau^*),
\ee
where the limit is taken with $\frac{\L}{n}=\mu$ held constant and
\be\label{eq:tau_MAP}
\tau^{*} = \sigma^2 + \mu\potargmin(\mu,\sigma^2).
\ee
We expect that a similar result can be shown via replica analysis for $B >1$ and general discrete priors.  Several arguments were put forward in \cite[Sec.~IV.A]{fengler2021sparcs} to suggest that such an extension is possible. 
\end{remark}

\begin{remark}\label{rem:diff}
The theoretical analysis in this paper (Theorems \ref{thm:amp_perf} to \ref{thm:sc_amp_asympB}) is similar to the state evolution and potential function analyses of SPARCs and random linear estimation \cite{barbier2016proof, barbier2017approximate, rush2020capacity, reeves2019thereplica, barbier2020mutual}.
The key technical differences are: 
(i) our coding scheme is a more general random linear framework than SPARCs as we allow the sections of the message vector to be drawn from a general discrete distribution over length $\B$ vectors (standard SPARCs use the specific distribution in Example \ref{ex:distr_unmod}), and
(ii) state evolution and potential function results are usually given in terms of the mean-squared error, whereas our results are given in terms of the UER, which requires additional technical steps, e.g., the sandwiching argument used to prove the second part of Theorem \ref{thm:amp_perf}.
\end{remark}

\subsection{Spatially Coupled Gaussian Matrices}

\begin{theorem}[Spatially Coupled Gaussian Matrices with AMP Decoding]\label{thm:threshold_sat}
Consider the linear model \eqref{eq:linear_regression} with a spatially coupled design matrix $\bA$ constructed using an $(\omega, \Lambda, \rho)$ base matrix, and the $L$ sections of the message vector $\bx$ i.i.d. $\sim p_{\bXsec}$.
Let $\hat{\bx}^t$ be the AMP hard-decision estimate of $\bx$ after iteration $t$ (see \eqref{eq:MAPdef}), 
and recall that $\btau^t\in\reals^{\Lc}$ is an output of the state evolution recursion \eqref{eq:sc_se_phi}--\eqref{eq:sc_se_psi} with $\Lc=\Lambda$ in this setting.

1) For any $(\omega,\Lambda,\rho)$ base matrix, each entry of $\btau^t \in \reals^ \Lc$  is non-increasing in $t$, and the $\sfc$th entry converges to a fixed point, denoted by  $\tau_\sfc^{\text{SC-FP}}$, for $\sfc\in[\Lc]$.

2) For any $\epsilon > 0$, there are constants $\omega_0 <\infty$, $\Lambda_0 < \infty$ and $\rho_0>0$ such that, for all $\omega > \omega_0$, $\Lambda > \Lambda_0$ and $0\leq\rho<\rho_0$,
the fixed points $\{\tau_\sfc^{\text{SC-FP}}\}_{\sfc\in[\Lc]}$ satisfy
\be\label{eq:sc_se_fp_tau}
\max_{\sfc\in[\Lc]} \tau_\sfc^{\text{SC-FP}}
\leq 
\overline{\tau}_{\vartheta}
\coloneqq \sigma^2 + \vartheta \mu (\max \potargmin(\vartheta \mu,\sigma^2) +\epsilon),
\ee
where $\vartheta = 1 +  \frac{(\omega-1)}{\Lambda}$, and the set of potential function minimizers $\potargmin(\vartheta \mu, \sigma^2)$ is defined in \eqref{eq:potential_func_argmin}.

3) Fix base matrix parameters $\omega>\omega_0$, $\Lambda>\Lambda_0$, and $0<\rho<\rho_0$. Fix $\delta>0$, and let $T$ denote the first iteration for which $\max_\sfc \tau_\sfc^t \leq \overline{\tau}_{\vartheta} + \delta$.
Then the UER of the AMP decoder after $T+1$ iterations satisfies the following almost surely:
\begin{align}\label{eq:sc_amp_se_convergence}
\lim_{\L\to\infty} 
\frac{1}{\L} \sum_{\ell=1}^\L \mathbbm{1} \big\{\hat{\bx}_\ell^{T+1} \neq \bx_\ell \big\} 
\stackrel{\text{a.s.}}{=} \frac{1}{\Lc} \sum_{\sfc=1}^{\Lc} P_e(\tau_\sfc^T)
\leq P_e(\overline{\tau}_{\vartheta} + \delta),
\end{align}
where the limit is taken with $\frac{\L}{n}=\mu$ held constant.
\end{theorem}
\begin{remark}[Threshold Saturation]\label{rmk:threshold_sat}
Theorem \ref{thm:threshold_sat} shows that the asymptotic UER achievable with a suitable spatially coupled Gaussian matrix and AMP decoding is  bounded by 
$P_e(\overline{\tau}_{\vartheta} +\delta)$.
%
As $\omega/\Lambda \to 0$, we have $\vartheta\to 1$.
Therefore, if $\potargmin(\mu,\sigma^2)$ defined in \eqref{eq:potential_func_argmin} is a singleton, (noting that $\epsilon$ in \eqref{eq:sc_se_fp_tau} can be arbitrarily small) we have
\be\label{eq:2tauMAP}
\lim_{\omega\to\infty} \lim_{\Lambda\to\infty} \overline{\tau}_{\vartheta} \to \tau^{*},
\ee
where $\tau^{*}$ is defined in \eqref{eq:tau_MAP}. 
Therefore, in the limit described in \eqref{eq:2tauMAP}, the asymptotic UER of the spatially coupled scheme with AMP decoding is bounded by $P_e(\tau^{*}+\delta)$ for any fixed $\delta>0$. This matches the (predicted) asymptotic UER  achieved by i.i.d. Gaussian matrices and SBS-MAP decoding (Remark \ref{rem:replica}).  This phenomenon, where the performance of message passing decoding in a spatially coupled system matches the optimal decoding performance in the corresponding uncoupled system, has been shown in other applications and is known as \emph{threshold saturation} \cite{kudekar2011threshold, kudekar2013spatially, donoho2013information, yedla2014asimple, kudekar2015wave, barbier2016proof}.
\end{remark}

\begin{IEEEproof}[Proof of Theorem \ref{thm:threshold_sat}]
1) Consider the spatially coupled state evolution \eqref{eq:sc_se_phi}--\eqref{eq:sc_se_psi} as a single line recursion in the vector $\boldsymbol{\gamma}^t\in\reals^{\Lr}$: for $\sfr \in [\Lr]$,
\be\label{eq:sc_se_single}
\gamma_\sfr^{t+1} =  \sum_{\sfc=1}^{\Lc} W_{\sfr \sfc} \, \mmse\bigg(\sum_{\sfr'=1}^{\Lr}W_{\sfr' \sfc} \, \frac{1}{\sigma^2 + \mu_{\text{inner}} \gamma_{\sfr'}^t}\bigg).
\ee
%
%
For any base matrix $\bW$ with non-negative entries, the result in \cite[Cor.~4.3]{barbier2016proof} 
shows that each entry of $\boldsymbol{\gamma}^t$ is non-increasing in $t$
and converges to a fixed point. We denote these fixed points by $\{\gamma_\sfr^{\text{SC-FP}}\}_{\sfr\in[\Lr]}$. 
The arguments used in \cite[Cor.~4.3]{barbier2016proof} are similar to those used in the proof of the first part of Theorem \ref{thm:amp_perf}.
The entries of the state evolution parameter $\btau^{t}$ given by
\begin{equation*}
\tau_\sfc^t = \bigg[\sum_{\sfr=1}^{\Lr}\frac{W_{\sfr \sfc}}{\sigma^2 + \mu_{\text{inner}}  \gamma_\sfr^t}\bigg]^{-1}, \quad \text{for } \sfc\in[\Lc],
\end{equation*}	
are non-decreasing in $\{\gamma_\sfr^{t}\}_{\sfr\in[\Lr]}$. Hence
we conclude that each entry of $\boldsymbol{\tau}^t$ is also non-increasing in $t$ and converges to a fixed point; these fixed points are denoted  $\{\tau_\sfc^{\text{SC-FP}}\}_{\sfc \in [\Lc]}$.
%


2) The result in \eqref{eq:sc_se_fp_tau} is obtained by using results from \cite{yedla2014asimple} that bound the fixed points of general coupled recursions. 
%
%
%
The uncoupled state evolution \eqref{eq:se} can be written as a single line recursion:
\be\label{eq:se_single2}
\psi^{t+1} =  \mmse \bigg(\frac{1}{\sigma^2 + \mu \psi^t}\bigg).
\ee
The uncoupled recursion in \eqref{eq:se_single2} and the coupled recursion in \eqref{eq:sc_se_single} correspond exactly to \cite[Eqs.~(27)-(28)]{yedla2014asimple} when $\mu$ of the uncoupled system is equal to $\mu_{\text{inner}}$ of the spatially coupled system  and $\bW$ is an $(\omega, \Lambda,\rho=0)$ base matrix.
(We will discuss the implications of $\rho$ being a small positive constant later.)
%
Using the same arguments as in \cite[Sec.~VI.E]{yedla2014asimple} and the vector I-MMSE relationship \cite[Thm.~2]{guo2005mutual}, we obtain the following result by applying \cite[Theorems 1 and 2]{yedla2014asimple}. 

For $\rho=0$ and any $\epsilon > 0$, there is an $\omega_0 <\infty$ and $\Lambda_0 < \infty$ such that, for all $\omega > \omega_0$ and $\Lambda > \Lambda_0$,
the fixed point of \eqref{eq:sc_se_single} satisfies
\be\label{eq:sc_se_fp_gamma}
\min \tilde{\mc{M}}(\mu_{\text{inner}}, \sigma^2) - \epsilon 
\leq
\max_{\sfr\in[\Lr]} \gamma_\sfr^{\text{SC-FP}}
\leq 
\max \tilde{\mc{M}}(\mu_{\text{inner}}, \sigma^2) +\epsilon,
\ee
where
\begin{align*}
\tilde{\mc{M}}(\mu, \sigma^2)
&= \argmin_{\psi\in[0,E]} \tilde{\mc{F}}(\mu, \sigma^2, \psi),\\
\tilde{\mc{F}}(\mu, \sigma^2, \psi) 
&= 2 \bigg\{ I\bigg(\bXsec; \sqrt{\frac{1}{\sigma^2 + \mu \psi}} \bXsec + \bZ\bigg) \nonumber\\
&\hspace{1em}- I\bigg(\bXsec; \sqrt{\frac{1}{\sigma^2}} \bXsec + \bZ\bigg) \nonumber\\
&\hspace{1em}+ \frac{1}{2\mu} \bigg[\ln\Big(1 + \frac{\mu\psi}{\sigma^2}\Big) - \frac{\mu\psi}{\sigma^2 + \mu\psi}\bigg] 
\bigg\}. \label{eq:potential_func_yedla}
\end{align*}
Here $\bXsec \sim p_{\bXsec}$ and $\bZ\in \reals^{\B}$ is a standard Gaussian vector independent of $\bXsec$.
Since $\tilde{\mc{F}}(\mu, \sigma^2, \psi)$ and the potential function $\pot(\mu, \sigma^2, \psi)$ defined in \eqref{eq:potential_func} are equivalent after removing constant scaling factors and terms that don't depend on $\psi$, their minimizers with respect to $\psi$ are identical.
Therefore, we can write \eqref{eq:sc_se_fp_gamma} as 
\be\label{eq:sc_se_fp_gamma2}
\min \potargmin(\mu_{\text{inner}},\sigma^2) - \epsilon 
 \leq
\max_{\sfr\in[\Lr]} \gamma_\sfr^{\text{SC-FP}}
\leq 
\max \potargmin(\mu_{\text{inner}},\sigma^2) +\epsilon,
\ee
where $\potargmin(\mu, \sigma^2)$ is the set of minimizers of $\mc{F}(\mu, \sigma^2, \psi)$.

Now we consider the effect of $\rho$ being a small positive constant on the fixed point of the state evolution. We study this scenario as $\rho$ needs to be lower bounded by a strictly positive constant for the AMP concentration result in \eqref{eq:sc_amp_se_convergence} to hold.
First, the $\mmse(s)$ function defined in \eqref{eq:mmse_func} is a smooth function of $s$ on $(0,\infty)$ \cite[Prop.~7]{guo2011estimation}.
%
Therefore, the right-hand-side of \eqref{eq:sc_se_single} is a smooth function of the entries of $\bW$.
%
%
Hence, the fixed point of the state evolution recursion \eqref{eq:sc_se_single} is a smooth function of $\rho$. For $\rho \ge 0$, denoting this fixed point by $\{ \gamma_{\sfr}^{\text{SC-FP}}(\rho) \}_{\sfr \in [\Lr]}$, and  letting
\[ \Delta(\rho) := \max_{\sfr \in [\Lr]} \,  \abs{ \gamma_{\sfr}^{\text{SC-FP}}(\rho) \, - \gamma_{\sfr}^{\text{SC-FP}}(0)},
\]
we  have $\Delta(\rho) \to 0$ as $\rho \to 0$.
Consequently, the result for $(\omega,\Lambda,\rho=0)$ base matrices in
\eqref{eq:sc_se_fp_gamma2} holds for $(\omega,\Lambda,\rho>0)$ base matrices with the deviation $\epsilon$  replaced by the slightly larger value $\epsilon + \Delta(\rho)$.
Equivalently, since $\epsilon >0$ is arbitrary and $\Delta(\rho)$ is a smooth function with $\Delta(0)=0$, there exists  $\rho_0>0$ such that for all $\rho < \rho_0$,
the result \eqref{eq:sc_se_fp_gamma2} holds for $(\omega,\Lambda,\rho>0)$ base matrices.


We now obtain \eqref{eq:sc_se_fp_tau} using \eqref{eq:sc_se_fp_gamma2}. 
For $\sfc\in[\Lc]$, we have
\begin{align}
\tau_\sfc^{\text{SC-FP}} 
&= \Bigg[\sum_{\sfr=1}^{\Lr}\frac{W_{\sfr \sfc}}{\sigma^2 + \mu_{\text{inner}}  \gamma_\sfr^{\text{SC-FP}} }\Bigg]^{-1}\nonumber\\
&\leq \Bigg[\frac{\sum_{\sfr=1}^{\Lr} W_{\sfr \sfc}}{\sigma^2 + \mu_{\text{inner}}  \max_{\sfr'\in[\Lr]} \gamma_{\sfr'}^{\text{SC-FP}} }\Bigg]^{-1}\nonumber\\[0.5em]
&\leq \sigma^2 + \mu_{\text{inner}}  
(\max \potargmin(\mu_{\text{inner}},\sigma^2) +\epsilon),\nonumber
\end{align}
where the last inequality is obtained using the $\sum_{\sfr=1}^{\Lr} W_{\sfr \sfc}=1$ constraint on base matrices, and the upper bound in \eqref{eq:sc_se_fp_gamma2}.
The result \eqref{eq:sc_se_fp_tau} follows by recalling from \eqref{eq:mu_inner} that $\mu_{\text{inner}} = \vartheta\mu$, where $\vartheta =1 + (\omega-1)/\Lambda$.

3) We now prove \eqref{eq:sc_amp_se_convergence}. 
Consider the input to the AMP hard-decision step in iteration $t+1$,
which we denote by $\bs^t = \bx^t + (\widetilde{\bS}^t \odot \bA)^*  \bz^t$ (see \eqref{eq:sc_amp}, \eqref{eq:MAPdef}).
For $\ell \in [L]$, we denote by $\boldsymbol{a}_\ell \in \reals^B$  the $\ell$th section of a vector $\boldsymbol{a} \in \reals^{LB}$.

The results in \cite{javanmard2013state} and  \cite{rush2020capacity} imply that for any pseudo-Lipschitz function $\xi: \reals^{B} \times \reals^{B} \to \reals$, the following holds almost surely:
\be \label{eq:PL2_res_sc}
\lim_{L \to \infty} \frac{1}{L} 
\sum_{\ell=1}^{\L} \xi(\bx_{\ell}, \bs^t_{\ell} ) 
= \frac{1}{\Lc} \sum_{\sfc=1}^{\Lc} \expec\{ \xi( \bXsec, \, \bS_{\tau_\sfc^t} ) \},
\ee
where the limit is taken with $\L/n= \mu$ held constant,
$\bXsec \sim p_{\bXsec}$ and $\bS_{\tau_\sfc^t}$ is given by \eqref{eq:single_user_ch}.  This result was proved in \cite{javanmard2013state} for the $B=1$ case, and extended in \cite{rush2020capacity} to the setting of SPARCs where the specific distribution $p_{\bXsec}$ given in Example \ref{ex:distr_unmod} (corresponding to random codebooks) is used. The proof for more general discrete distributions is essentially the same.  Then, following the same steps as \eqref{eq:MAPdecregion}--\eqref{eq:MAP_Pe_conv}
(using \eqref{eq:PL2_res_sc} instead of \eqref{eq:PL2_res} in \eqref{eq:epm_limits}) gives the desired result.
\end{IEEEproof}

\subsection{Numerical Results}\label{sec:sc_gauss_sim}


\begin{figure*}[!t]
\subfloat[2 bits ($\B=2^2$)]{
	\includegraphics[width=0.48\textwidth]{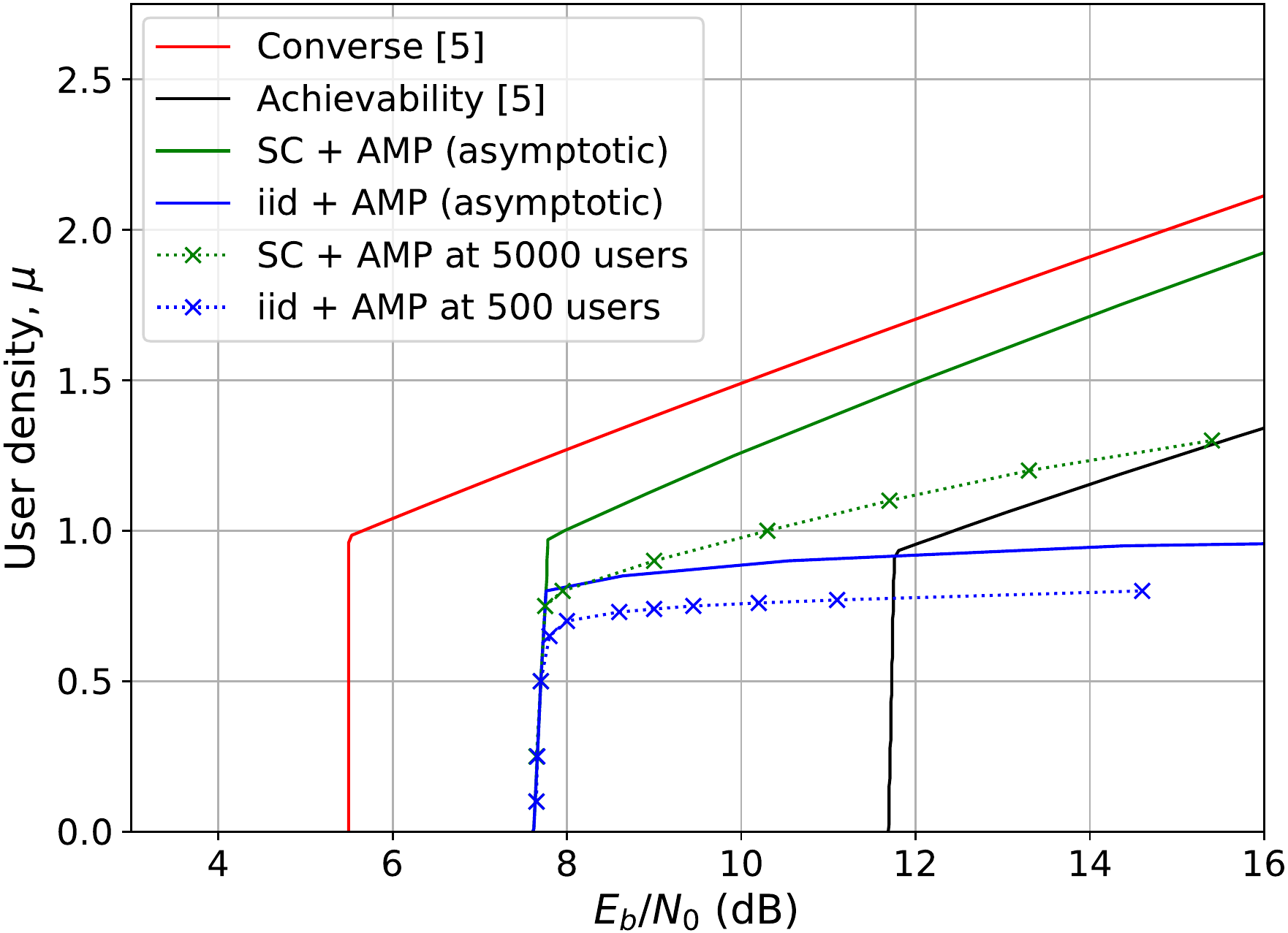}%
	\label{fig:gmac_amp_scamp_sim_k2}}
\hfill
\subfloat[8 bits ($\B=2^8$)]{
	\includegraphics[width=0.48\textwidth]{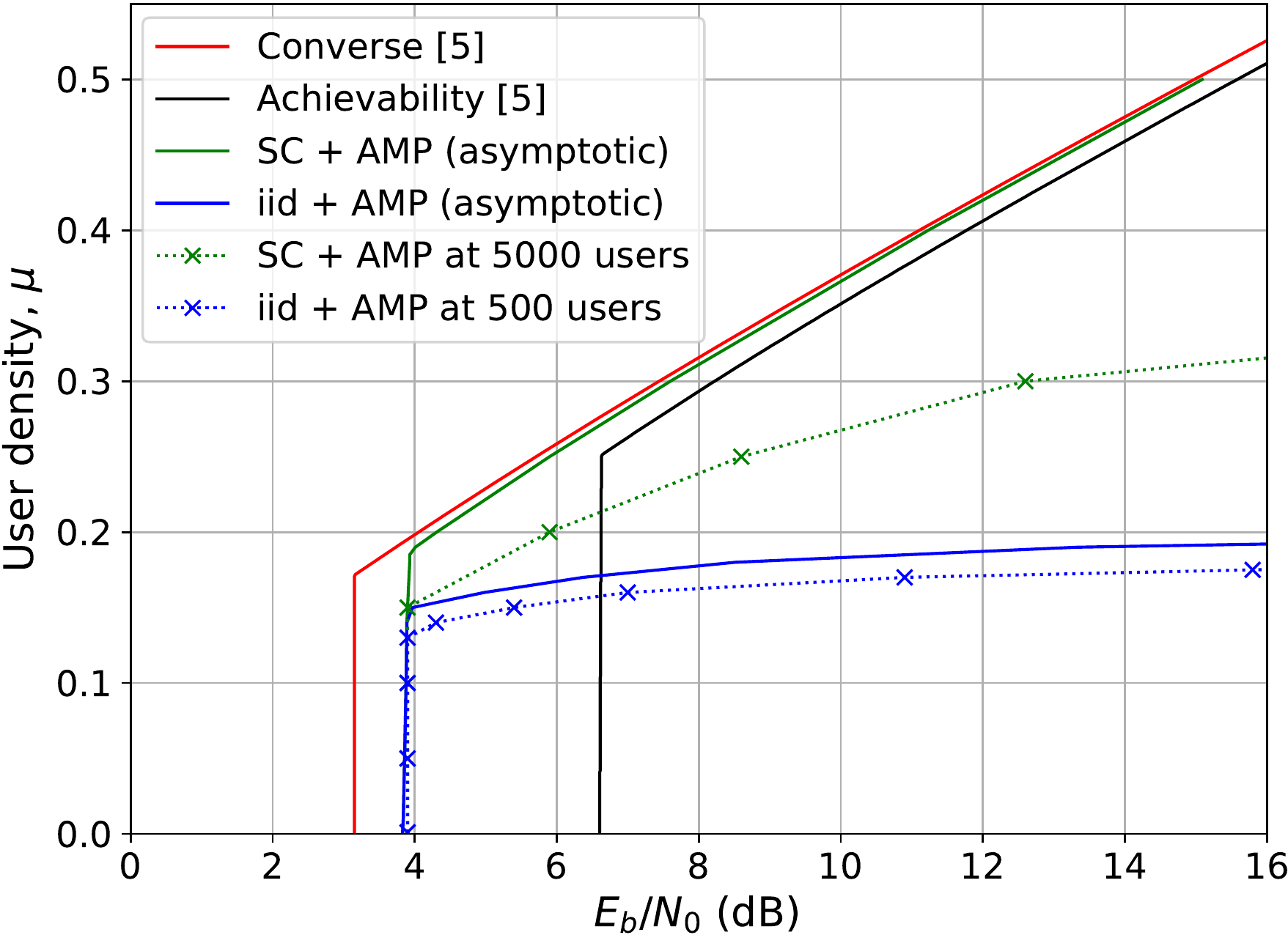}%
	\label{fig:gmac_amp_scamp_sim_k8}}
\caption{Achievable regions of i.i.d. and spatially coupled Gaussian codebooks with AMP decoding, when the user payload is either 2 bits or 8 bits, and the the maximum tolerated expected (or average) UER is $10^{-3}$.
}
\label{fig:gmac_amp_scamp_sim}
\end{figure*}


\begin{table}[!t]
\renewcommand{\arraystretch}{.9}
\caption{Optimized coupling width values used in Figs.\ \ref{fig:gmac_amp_scamp_sim_k2} and \ref{fig:gmac_amp_scamp_sim_k8}.}
\label{table:scamp_sim_omega}
\centering
 \begin{tabular}{| c || c c c c c |} 
 \hline
& \multicolumn{5}{c|}{Fig.\ \ref{fig:gmac_amp_scamp_sim_k2}} \\
 \hline
  $\mu$ & 0.9 & 1.00 & 1.1 & 1.2 & 1.3  \\ 
 \hline
  $\omega$ & 5 & 5 & 6 & 6 & 7 \\
 \hline
 \hline
 & \multicolumn{5}{c|}{Fig.\ \ref{fig:gmac_amp_scamp_sim_k8}} \\
 \hline
 $\mu$ & 0.15 & 0.20 & 0.25 & 0.30 & 0.33 \\
  \hline
 $\omega$ & 5 & 5 & 6 & 11 & 14 \\
 \hline
\end{tabular}
\end{table}

Theorems \ref{thm:amp_perf} and \ref{thm:threshold_sat} together with Remark \ref{rmk:threshold_sat} give us the asymptotic UER achieved by random linear coding and AMP decoding when i.i.d. and spatially coupled Gaussian design matrices are used.
These results are given in terms of the largest stationary point and global minimum of the potential function defined in \eqref{eq:potential_func}.
In this section, we numerically evaluate these results to understand the achievable regions of random linear coding schemes with AMP decoding.

The solid blue and green curves in Figs. \ref{fig:gmac_amp_scamp_sim_k2} and \ref{fig:gmac_amp_scamp_sim_k8} plot the asymptotic achievable region of AMP decoding 
when the Gaussian design matrix $\bA$ is either \iid\ or spatially coupled, and the sections of the message vector $\bx$ are drawn \iid\ from $\px$ (see Example \ref{ex:distr_unmod}).
Specifically, for a list of user densities $\mu$, we plot the minimum $E_b/N_0$ required by the coding schemes to achieve a UER of less than $10^{-3}$, when the user payload is $2$ or $8$ bits.
Recall that the asymptotic UER achieved by the two schemes are of the form $P_e(\tau)$, 
where $P_e(\cdot)$ is defined in \eqref{eq:Pe_unmod} and $\tau$ is specified by \eqref{eq:se_fp_pot_tau} for the i.i.d. scheme and by \eqref{eq:tau_MAP} for the spatially coupled scheme. Furthermore, recall that $E_b$ and $N_0$ are related to $E$ and $\sigma^2$ using $E = E_b \log_2 \B$ and $N_0 = 2\sigma^2$.

%
In both Figs. \ref{fig:gmac_amp_scamp_sim_k2} and \ref{fig:gmac_amp_scamp_sim_k8}, the asymptotic achievable region of spatially coupled Gaussian codebooks with efficient AMP decoding (solid green) is strictly larger than the achievability bound in \cite{zadik2019isit} (solid black), which is based on i.i.d. Gaussian codebooks and ML decoding.
Moreover, in Fig. \ref{fig:gmac_amp_scamp_sim_k8} where the user payload is 8 bits, it nearly matches the converse bound in \cite{zadik2019isit} (solid red) for $\mu \geq 0.2$.  The converse and achievability bounds in \cite{zadik2019isit} are given in terms of the expected UER, which is also known as the per-user probability of error. See the end of this section for more details on the converse bound.

At low user densities ($\mu \leq 0.80$ in Fig. \ref{fig:gmac_amp_scamp_sim_k2} and $\mu \leq 0.15$ in Fig. \ref{fig:gmac_amp_scamp_sim_k8}), we observe that the minimum $E_b/N_0$ required by the i.i.d. and spatially coupled coding schemes is the same; this is because the global minimum of the potential function coincides with its largest stationary point. However, the gap between the achievable regions of the two schemes increases sharply for larger $\mu$. Furthermore, the shape of the solid blue curve for large $E_b/N_0$ suggests that it might be impossible to achieve $\text{UER} \leq 10^{-3}$ with i.i.d. Gaussian codebooks and AMP decoding above a certain user density ($\mu \approx 1.0$ in Fig. \ref{fig:gmac_amp_scamp_sim_k2} and $\mu \approx 0.2$ in Fig. \ref{fig:gmac_amp_scamp_sim_k8}).

The blue and green dotted lines with crosses in Figs. \ref{fig:gmac_amp_scamp_sim_k2} and \ref{fig:gmac_amp_scamp_sim_k8} show the simulated performance of the i.i.d. and spatially coupled coding schemes with 500 and 5000 users, respectively.
For a list of user densities $\mu$, the crosses show the minimum $E_b/N_0$ at which the coding scheme achieves an average UER of less than $10^{-3}$ (averaged over many independent trials).
Recall the construction details of spatially coupled matrices in Section \ref{sec:sc_coding} and the description of AMP decoding in Section \ref{sec:scamp}.
Discrete Cosine Transform (DCT) based design matrices were used instead of Gaussian ones to reduce decoding complexity and memory usage---the error rates obtained by the two methods are similar for large matrix sizes.
See \cite{rush2017capacity, barbier2017approximate}, \cite[Sec.~2.5.1]{hsieh2021thesis} for details on the DCT implementation.%
\footnote{Although the cited works use Hadamard based design matrices and the Fast Walsh-Hadamard Transform in their simulations, our DCT construction is essentially the same.}
%
%
The simulations for the spatially coupled coding scheme used $(\omega, \Lambda, \rho)$ base matrices with $\Lambda=50, \rho=0$.  For each user density $\mu$, we calculated the average UER obtained via simulations for a range of coupling widths $\omega$ and picked the best one. Table I gives the optimized $\omega$ for each $\mu$, with $\Lambda=50, \rho=0$ fixed.


%

We observe that for both i.i.d. and spatially coupled coding schemes, the finite user and asymptotic curves match at low user densities (the near-vertical part of the curve). 
For the i.i.d. coding scheme (blue), although a gap between the two curves appears above a certain user density threshold ($\mu = 0.65$ in Fig. \ref{fig:gmac_amp_scamp_sim_k2} and $\mu = 0.13$ in Fig. \ref{fig:gmac_amp_scamp_sim_k8}), their overall shape remains similar. 
For the spatially coupled scheme (green), the gap between the asymptotic and finite user curves 
increases with $\mu$. This gap is a finite length effect, due to the relatively small values of base matrix parameters.

Table \ref{table:scamp_sim_omega} shows the values of the optimized coupling widths used in Figs. \ref{fig:gmac_amp_scamp_sim_k2} and \ref{fig:gmac_amp_scamp_sim_k8} for user densities 
above the near-vertical parts of each curve.
%
We observe that the optimal coupling width increases with the user density.
%
%
At lower user densities, a range of coupling widths (including the uncoupled case $\omega=1$) achieve similar average UERs.

The solid red curves in Figs.  \ref{fig:gmac_amp_scamp_sim_k2} and \ref{fig:gmac_amp_scamp_sim_k8} show the following converse bound  from \cite{zadik2019isit} on the minimum $E_b/N_0$ to achieve  expected UER within a given $\epsilon>0$ 
(for fixed user density $\mu$ and user payload $\log_2\M$):
\be\label{eq:converse_min_energy}
\begin{split}
\frac{E_b}{N_0} \geq \max\Bigg\{ &\frac{\left[Q^{-1}\left(\frac{1}{\M}\right) - Q^{-1}(1-\epsilon)\right]^2}{2 \log_2 \M}, \, \\
& \frac{2^{2\mu\left[\log_2\M - \epsilon \log_2 (\M-1) - H_b(\epsilon)\right]} - 1}{2\mu \log_2\M} \Bigg\},
\end{split}
\ee
where $H_b(\cdot)$ is the binary entropy function.
%


\section{Large User Payloads}\label{sec:large_M}

When coding with random codebooks, i.e., when the sections of $\bx$ are drawn i.i.d. from $\px$ (see Example \ref{ex:distr_unmod}), the size $B$ of each section in $\bx$ increases exponentially with the user payload (which is $\log_2 \M= \log_2 \B$ bits).
For very large $\B$ it is computationally infeasible to evaluate the potential function \eqref{eq:potential_func}. The potential function is needed to compute the asymptotic UER bounds in Theorems \ref{thm:amp_perf} and \ref{thm:threshold_sat} (see \eqref{eq:amp_se_convergence} and \eqref{eq:sc_amp_se_convergence}). In this section we bound the asymptotic UER achieved by i.i.d. and spatially coupled Gaussian codebooks with AMP decoding when the user payload is large.
Furthermore, in Section \ref{sec:large_M_implement} we discuss simple ways to reduce the decoding complexity at larger user payloads.
Both results in this section (Theorems \ref{thm:amp_asympB} and \ref{thm:sc_amp_asympB}) use the following lemma.

\begin{lemma}[Asymptotic UER Bound]\label{lem:uer_mse}
Consider the setting of either Theorem \ref{thm:amp_perf} or \ref{thm:threshold_sat}, and take the distribution $p_{\bXsec}$ to be $\px$.
Let $\hat{\bx}^t$ be the AMP hard-decision estimate of $\bx$ after iteration $t$, 
and recall that $\bpsi^t \in \reals^{\Lc}$ is an output of the state evolution \eqref{eq:sc_se_phi}--\eqref{eq:sc_se_psi}.
Then we have that the following limit exists almost surely and satisfies:
\begin{align}
\label{eq:uer_mse_bounds}
\lim_{\L\to\infty}
\frac{1}{\L} \sum_{\ell=1}^\L \mathbbm{1} \big\{\hat{\bx}_\ell^t \neq \bx_\ell \big\} 
\leq 
\frac{4}{\Lc} \sum_{\sfc=1}^{\Lc} \frac{\psi_\sfc^t}{E},
\end{align}
where the limit is taken with $\frac{\L}{n}=\mu$ held constant.
Recall that $\Lc=1$ when the design matrix $\bA$ has i.i.d. Gaussian entries $\mc{N}(0,\frac{1}{n})$.
\end{lemma}

The proof of Lemma \ref{lem:uer_mse} is given in Appendix \ref{app:proof_lem_uer_mse}.

\subsection{IID Gaussian Codebooks}

\begin{theorem}[AMP Decoding of IID Gaussian Codebooks at Large User Payloads]\label{thm:amp_asympB}
Consider the setting of Theorem \ref{thm:amp_perf} and take the distribution $p_{\bXsec}$ to be $\px$.
The UER of the AMP decoder after its first iteration exhibits the following phase transition for sufficiently large payloads $\log_2 \B$.

1) For any $\delta\in(0, \frac{1}{2})$, let $f_{\B,\delta}\coloneqq \frac{\B^{-k\delta^2}}{\delta\sqrt{\ln \B}}$ where $k$ is a positive constant.
If
\be \label{eq:low_mu_regime}
\mu \log_2 \B < \frac{1}{2}  \left(\frac{1}{(1+\frac{\delta}{2})\ln 2} - \frac{1}{E_b/N_0}\right), 
\ee
then $\lim_{\L \to\infty} \frac{1}{\L} \sum_{\ell=1}^\L \indic\big\{\hat{\bx}_\ell^1 \neq \bx_\ell \big\} \leq  4 f_{\B,\delta}$.

2) For any $\tilde{\delta}\in(0, 1)$ and $\delta_2 \in (0, \sqrt{2} - \sqrt{2 - \tilde{\delta}})$,
let $g_{\B,\tilde{\delta}}\coloneqq \B^{-k_1{\tilde{\delta}^2}}$ where $k_1$ is a positive constant and
$h_{\B,\delta_2} \coloneqq  \frac{B^{-\delta^2_2/2}}{\delta_2 \sqrt{ \ln \B} } + B^{-\delta_2^2}$.
If 
\be \label{eq:hi_mu_regime}
\mu \log_2 \B > \frac{1}{2(1-g_{\B,\tilde{\delta}})} \left(\frac{1}{(1-\frac{\tilde{\delta}}{2})\ln 2} - \frac{1}{E_b/N_0} \right),
\ee
then 
$\lim_{\L \to\infty} \frac{1}{\L} \sum_{\ell=1}^\L \indic\big\{\hat{\bx}_\ell^t \neq \bx_\ell \big\} \geq 1 - h_{\B,\delta_2}$ 
for all $t \geq 1$.

In both statements, the limits exist and are taken with $\frac{\L}{n}=\mu$ held constant.
\end{theorem}

\begin{remark}\label{rmk:amp_asympB_spectraleff} 
From \eqref{eq:hi_mu_regime}, we see that for any fixed values of $\mu$ and $\frac{E_b}{N_0}$, the UER of AMP decoding is lower bounded by a value that approaches $1$ with growing $B$. Therefore, the interesting regime for large user payloads is when the spectral efficiency
\be
S := \mu \log_2 \B = \frac{\L \log_2 \B}{n} \quad \text{bits/transmission},
\ee
is of constant order. (The spectral efficiency is the total number of bits  transmitted by all the users per channel use.)
Theorem \ref{thm:amp_asympB} can be extended (using analysis similar to  \cite{rush2017capacity,rush2020capacity}) to this asymptotic regime where $\L, n, \log_2\B$ all tend to infinity with the spectral efficiency held constant. 
In this case, the UER of the AMP decoder 
exhibits the following phase transition in this large system limit almost surely:
\begin{align}\label{}
\lim_{\L,\B, n\to\infty} 
\frac{1}{\L} \sum_{\ell=1}^\L \mathbbm{1} \big\{\hat{\bx}_\ell^1 \neq \bx_\ell \big\} 
\, = \,
\begin{cases}\label{eq:amp_asympB_spectraleff}
	0  &\text{if } S < S_{\text{AMP}},\\
	1  &\text{otherwise},
\end{cases}
\end{align}
where $S_{\text{AMP}}$ is defined as
\be\label{eq:S_BP}
S_{\text{AMP}} \coloneqq \frac{1}{2} \bigg(\frac{1}{\ln 2} - \frac{1}{E_b/N_0}\bigg).
\ee
From \eqref{eq:amp_asympB_spectraleff} and \eqref{eq:S_BP}, we see that positive spectral efficiencies are achievable in this large system setting using i.i.d. Gaussian codebooks and AMP decoding if and only if $\frac{E_b}{N_0} > \ln 2$.
\end{remark}

\begin{IEEEproof}[Proof of Theorem \ref{thm:amp_asympB}]
Recalling the definition of the state evolution parameters $\tau^t$ and $\psi^t$ from \eqref{eq:se}, let
\be\label{eq:se_nu}
\nu^t \coloneqq \frac{E}{\tau^t \ln \B}
= \frac{E}{(\sigma^2 + \mu \psi^t) \ln B}.
\ee

From \cite[Lem.~4.1]{rush2020capacity} we know that for sufficiently large $\B$ and any $\delta\in(0, \frac{1}{2})$, $\tilde{\delta}\in(0, 1)$, we have
\be\label{eq:largeM_psi_lb_ub_nu}
(1 - g_{\B,\tilde{\delta}})\mathbbm{1}\{\nu^t < 2 - \tilde{\delta}\}
< 
\frac{\psi^{t+1}}{E}
\leq 
1 - (1 - f_{\B,\delta})  \mathbbm{1}\{\nu^t > 2 + \delta\},
\ee
where $f_{\B,\delta}$ and $g_{\B,\tilde{\delta}}$ are defined in the theorem statement.
We now prove the two parts of the theorem in sequence.

1) 
Using \eqref{eq:se_nu} in the upper bound of \eqref{eq:largeM_psi_lb_ub_nu} and recalling that  $E = E_b \log_2 B$, $\sigma^2=N_0/2$, we obtain
%
%
\be\label{eq:largeM_psi_ub}
\frac{\psi^{t+1}}{E} \leq  f_{\B,\delta}  
\quad \text{if} \quad \frac{\psi^t}{E} < \frac{\frac{2}{2+\delta} - \frac{\ln 2}{E_b/N_0}}{2\mu \ln \B}.
\ee

By substituting the initial condition $\psi^0=E$
into \eqref{eq:largeM_psi_ub}, 
we have that $\frac{\psi^1}{E} \leq f_{\B,\delta}$ under the condition in \eqref{eq:low_mu_regime}.
Furthermore, from Lemma \ref{lem:uer_mse} we know that the asymptotic UER after iteration $t = 1$ satisfies
\begin{align}
\lim_{\L\to\infty} \frac{1}{\L} \sum_{\ell=1}^\L \mathbbm{1} \big\{\hat{\bx}_\ell^1 \neq \bx_\ell \big\} 
\leq \frac{4\psi^1}{E} 
\leq 4 f_{\B,\delta}. \label{eq:largeM_uer_bounds}
\end{align}


2) 
We prove the second statement of the theorem by first showing that under \eqref{eq:hi_mu_regime}, 
\be\label{eq:Etau_bound}
\frac{E}{\tau^t} < (2-\tilde{\delta}) \ln \B \quad \text{for all } t \geq 0.
\ee
%

For $t=0$, we have $\tau^0 = \sigma^2 + \mu E$, and the condition in \eqref{eq:hi_mu_regime} ensures that $\frac{E}{\tau^0} < (2-\tilde{\delta}) \ln \B$.

Assume towards induction that $\frac{E}{\tau^t} < (2-\tilde{\delta}) \ln \B$ for some $t \geq 0$. Then we have
\be\label{eq:induc_1}
\frac{E}{\tau^{t+1}} 
= \frac{E}{\sigma^2 + \mu \psi^{t+1}}
< \frac{1}{(\sigma^2/E) + \mu (1-g_{\B,\tilde{\delta}})},
\ee
where the inequality is obtained from \eqref{eq:largeM_psi_lb_ub_nu}  and noting that $\nu^t = \frac{E}{\tau^t \ln \B} < 2- \tilde{\delta}$. We further bound the right side of \eqref{eq:induc_1} by  using the condition in \eqref{eq:hi_mu_regime} along with $E = E_b \log_2 B$ and $\sigma^2=N_0/2$ to obtain:
\be
\begin{split}
\frac{E}{\tau^{t+1}} 
&< \frac{\ln\B}{\frac{\ln 2}{2}\frac{1}{E_b/N_0} + \mu \log_2 \B (1-g_{\B,\tilde{\delta}}) \ln 2} \\
&< \frac{\ln\B}{\frac{\ln 2}{2}\frac{1}{E_b/N_0} + \frac{\ln 2}{2} (\frac{1}{(1-\tilde{\delta}/2)\ln 2} - \frac{1}{E_b/N_0} )} \\
%
%
&=(2-\tilde{\delta})\ln\B.
\end{split}
\ee
This shows that \eqref{eq:Etau_bound} holds for all $t \ge 0$.

Using \eqref{eq:Etau_bound}, we now prove that 
$\lim_{\L \to\infty} \frac{1}{\L} \sum_{\ell=1}^\L \indic\big\{\hat{\bx}_\ell^t \neq \bx_\ell \big\} \geq 1 - h_{\B,\delta_2}$ 
for all $t \geq 1$.
Using \eqref{eq:amp_se_convergence} and \eqref{eq:Pe_unmod} (noting that the first equality in \eqref{eq:amp_se_convergence} holds for $t\geq0$), we have
\be \label{eq:Pe_result}
\begin{split}
&\lim_{\L \to \infty}  \frac{1}{\L} \sum_{\ell=1}^\L \mathbbm{1} \big\{\hat{\bx}_\ell^{t+1} \neq \bx_\ell \big\} \\
&= P_e(\tau^t)
= 1 - \expec\left[\Phi\left(\sqrt{E/\tau^t} + Z\right)^{\B-1}\right],
\end{split}
\ee
where $Z \sim \mathcal{N}(0,1)$ and $\Phi(\cdot)$ is the distribution function of the standard normal. This holds specifically when $\bXsec\sim \px$. Thus to show the main result we will prove
\be
\label{eq:toprove1}
\expec\left[\Phi\left(\sqrt{E/\tau^t} + Z\right)^{\B-1}\right] \leq h_{\B, \delta_2} \,,
\ee
noting that $\frac{E}{\tau^t} < (2 - \tilde{\delta}) \ln \B$ as shown above.
We will use the following bounds on the tail probability of the standard normal $\Phi^c(u) \coloneqq \prob(Z > u)$ \cite{qfunc}.
For all $u > 0$,
\be
\sqrt{ \frac{2}{\pi}} \left(\frac{1}{u+\sqrt{u^2 + 4}}\right) e^{-u^2/2} 
< 
\Phi^c(u) 
< \frac{1}{\sqrt{2 \pi}} \left(\frac{1}{u}\right) e^{-u^2/2}.
\label{eq:Gaussian1}
\ee

For arbitrary $\delta_2 \in (0, \sqrt{2} - \sqrt{2 - \tilde{\delta}})$,
we have the bound
\be
\begin{split}
\label{eq:to_bound1}
& \expec\left[\Phi\left(\sqrt{E/\tau^t} + Z\right)^{\B-1}\right] \\
&\leq \prob\left( |Z| > \delta_2 \sqrt{ \ln \B}  \right) \\
&\quad + \expec\left[\Phi\left(\sqrt{E/\tau^t} + Z\right)^{\B-1} \Big \lvert \, |Z| \leq \delta_2\sqrt{ \ln \B}  \right].
\end{split}
\ee
Label the two terms on the right side of \eqref{eq:to_bound1} as $T_1$ and $T_2$. Using \eqref{eq:Gaussian1}, 
for the first term $T_1$ we have:
\begin{align} \label{eq:T1_bound1}
T_1 
= 2  \Phi^c\left(\delta_2 \sqrt{ \ln \B} \right) 
&\leq \sqrt{\frac{2}{ \pi}} \left(\frac{1}{ \delta_2 \sqrt{ \ln \B} }\right) e^{-(\delta_2^2 \ln \B)/2} \nonumber\\
&\leq \frac{B^{-\delta^2_2/2}}{\delta_2 \sqrt{ \ln \B} }.
\end{align}

To bound $T_2$,  first notice that  conditional on $|Z| \leq \delta_2 \sqrt{ \ln \B}$ and $\frac{E}{\tau^t} < (2 - \tilde{\delta}) \ln \B$,
we have 
\be
\label{eq:ab_bound} 
0  \leq \left \lvert\sqrt{E/\tau^t} + Z\right \lvert 
\leq (\sqrt{2 - \tilde{\delta}} + \delta_2) \sqrt{\ln \B}.
\ee
Using this and the substitution $\delta_3 \coloneqq \sqrt{2 - \tilde{\delta}} + \delta_2$, we can bound $T_2$ as follows:
\begin{align}
&T_2 
\stackrel{\text{(i)}}{\leq}  \expec\left[\left[1 - \Phi^c\left( \delta_3 \sqrt{\ln \B} \right)\right]^{\B-1}  \right]  \nonumber\\
&\stackrel{\text{(ii)}}{\leq}  \expec\left[\exp\left\{ - (\B-1)\Phi^c\left( \delta_3 \sqrt{\ln \B} \right)\right\}  \right]  \nonumber\\
&\stackrel{\text{(iii)}}{\leq}  \exp \left\{  \frac{-\sqrt{2}(\B-1)e^{ -  \delta_3^2 \ln \B/2 }}{\sqrt{\pi}\Big(\delta_3 \sqrt{\ln \B} + \sqrt{\delta_3^2 \ln \B + 4}\Big)}  \right\}   \nonumber\\
&\stackrel{\text{(iv)}}{\leq} \exp \left\{ \frac{-\sqrt{2}\B^{1 -  \delta_3^2/2 }}{5\sqrt{\pi \ln \B}}  \right\},
 \label{eq:T2_bound1}
\end{align}
where the labelled steps are obtained as follows:
(i) using $\Phi(u) \leq \Phi(|u|)$, 
the second inequality in \eqref{eq:ab_bound}, and $\Phi(u) = 1-\Phi^c (u)$;
(ii) using the bound $(1-x) \leq e^{-x}$ when $x > 0$;
(iii) using \eqref{eq:Gaussian1};
and
in (iv) we use $\B-1 > (4/5) \B$ when $\B > 5$,
and $\delta_3 = \sqrt{2 - \tilde{\delta}} + \delta_2 \leq \sqrt{2}$ when $\delta_2 \in (0, \sqrt{2} - \sqrt{2 - \tilde{\delta}})$, so that for $\ln \B \geq 4$
we have $\delta_3 \sqrt{\ln \B} + \sqrt{\delta_3^2 \ln \B + 4} 
 <  4 \sqrt{\ln \B}$.

Now we notice that \eqref{eq:T2_bound1} implies $T_2 \leq B^{-\delta_2^2}$ since
$ - \frac{\sqrt{2} \B^{1 -  (\sqrt{2 - \tilde{\delta}} + \delta_2)^2/2 }}{5\sqrt{\pi \ln \B}} 
\leq  -  \delta_2^2 \ln \B$
for large enough $\B$  (because $1 - (\sqrt{2 - \tilde{\delta}} + \delta_2)^2/2 > 0$). Finally, using \eqref{eq:to_bound1} and combining the bounds on $T_1$ and $T_2$, we conclude that 
$\expec [\Phi (\sqrt{E/\tau^t} + Z)^{\B-1}] \le h_{B, \delta_2}$.
\end{IEEEproof}

\subsection{Spatially Coupled Gaussian Codebooks}

From Remark \ref{rmk:amp_asympB_spectraleff}, we see that for large user payloads and spectral efficiencies less than $S_{\text{AMP}}$, one does not require spatial coupling for reliable AMP decoding.
The following result shows that any spectral efficiency above $S_{\text{AMP}}$ and below the converse can be achieved using spatially coupled Gaussian codebooks and AMP decoding.
The converse is discussed in Remark \ref{rmk:sc_param_choice}.

\begin{theorem}[AMP Decoding of Spatially Coupled Gaussian Codebooks at Large User Payloads]\label{thm:sc_amp_asympB}
Consider the setting of Theorem \ref{thm:threshold_sat} and take the distribution $p_{\bXsec}$ to be $\px$. Let
 $\vartheta = 1 + \frac{\omega-1}{\Lambda}$, $\mu=\frac{\L}{n}$, $\SNR = \frac{2E_b}{N_0} \mu \log_2 \B$, and define
\begin{align}
\Delta &\coloneqq \frac{1}{2\vartheta} \ln (1 + \vartheta \, \SNR) - \mu \ln \B, \label{eq:sc_amp_asympB_Delta}\\
\omega^* &\coloneqq \Big(\frac{\vartheta\, \SNR^2}{1+\vartheta\,\SNR} \Big) \frac{1}{\Delta},\label{eq:sc_amp_asympB_omega}\\
\rho^* &\coloneqq \min\Big\{ \frac{\Delta}{3 \, \SNR}, \frac{1}{2}\Big\}.
\end{align}
Let $\delta$ be an arbitrary constant in $(0,  \min\{\frac{\Delta}{2\mu \ln \B}, \frac{1}{2}\})$  and
$S_{\text{opt}}>0$ be the solution to
\be\label{eq:sc_amp_asympB_alpha}
S_{\text{opt}} = \frac{1}{2}\log_2 \bigg(1+ S_{\text{opt}} \, \frac{2 E_b}{N_0}\bigg).
\ee

1) If the spectral efficiency satisfies
\be\label{eq:sc_amp_asympB_mu_cond}
\frac{1}{\vartheta} S_{\text{AMP}}
\, \leq \, \mu \log_2 \B \, < \,
\frac{1}{\vartheta} S_{\text{opt}},
\ee
and the base matrix parameters satsify
$\omega > \omega^*$ and $0< \rho \leq \rho^*$,
then, for $t\geq1$ and 
$\sfc \leq \max \{ \frac{\omega t}{\omega^*}, \lceil \frac{\Lambda}{2}\rceil \}$, 
we have
\be\label{eq:}
\psi_\sfc^t = \psi_{\Lambda-\sfc+1}^t\leq E \, f_{\B,\delta}
\ee
for sufficiently large $\B$, where $E = E_b \log_2B $, $f_{\B,\delta}\coloneqq \frac{\B^{-k\delta^2}}{\delta\sqrt{\ln \B}}$ 
and $k$ is a positive constant.

2) Let $T$ denote the first iteration for which $\max_\sfc \psi_\sfc^t \leq E \, f_{\B,\delta}$. Then we have
\be
T \, \leq \, \bigg\lceil \frac{\Lambda \omega^\star}{2 \omega} \bigg\rceil,
\ee
and the UER of the AMP decoder after $T$ iterations satisfies the following almost surely:
\begin{align}
%
\lim_{\L\to\infty} 
\frac{1}{\L} \sum_{\ell=1}^\L \mathbbm{1} \big\{\hat{\bx}_\ell^T \neq \bx_\ell \big\}
\leq 4 f_{\B,\delta},\label{eq:largeM_scamp_uer_bound}
\end{align}
where the limit is taken with $\frac{\L}{n}=\mu$ held constant.
\end{theorem}
\begin{IEEEproof}
The first part of Theorem \ref{thm:sc_amp_asympB} is a direct application of the state evolution analysis of spatially coupled SPARCs for channel coding over the (single-user) AWGN channel \cite[Prop.~4.1]{rush2020capacity}.
The main change of variables required is that the 
signal-to-noise ratio in the AWGN channel is replaced by
$\SNR = \frac{L (E/n)}{\sigma^2} = \frac{2E_b}{N_0} \mu \log_2 \B$. Another change is that the AWGN rate $R = \frac{\L\ln \B}{n}$ in \cite{rush2020capacity} is replaced by $\mu \ln \B$.
The result in \eqref{eq:largeM_scamp_uer_bound} is a direct application of Lemma \ref{lem:uer_mse}.
\end{IEEEproof}

\begin{remark}
A positive solution to \eqref{eq:sc_amp_asympB_alpha} exists if and only if $\frac{E_b}{N_0} > \ln 2$.
\end{remark}

\begin{remark}[Parameter Choice]\label{rmk:sc_param_choice}
Consider spectral efficiency $S=\mu \log_2 \B$ bits/transmission. 
For any spectral efficiency $S < S_{\text{opt}}$, 
or equivalently any $\frac{E_b}{N_0} > \frac{2^{2 S}-1}{2 S}$ (which matches the converse bound in \cite{zadik2019isit} with $\B \to \infty$ and the target expected UER $\epsilon \to 0$), 
we can choose design parameters as follows to guarantee that the AMP decoder achieves a small UER at large payloads.

1) If $S<S_{\text{AMP}}$, or equivalently $\frac{E_b}{N_0} > (\frac{1}{\ln 2} - 2S)^{-1}$ for $S < \frac{1}{2\ln 2}$, we can use i.i.d. Gaussian codebooks. Theorem \ref{thm:amp_asympB} then guarantees that the UER is bounded by a small constant at large payloads.
 
2) If $S_{\text{AMP}} \leq S < S_{\text{opt}}$, we can use spatially coupled codebooks with base matrix parameters $\omega$ and $\Lambda$ chosen as follows to satisfy the conditions of Theorem \ref{thm:sc_amp_asympB}.
Letting $\vartheta_0 = S_{\text{opt}}/S$, first choose $\omega > \omega^\star (\vartheta_0)$ (defined as in \eqref{eq:sc_amp_asympB_omega} with $\vartheta$ replaced by $\vartheta_0$).
Then  choose $\Lambda$ large enough that $\vartheta = 1 + \frac{\omega-1}{\Lambda} \leq \vartheta_0$. 
This ensures that $S < S_{\text{opt}}/\vartheta$ and $\omega > \omega^\star(\vartheta)$.
Theorem \ref{thm:sc_amp_asympB} then guarantees that the UER is bounded by a small constant at large payloads.
\end{remark}

\begin{remark}\label{rmk:sc_amp_asympB_spectraleff}
Theorem \ref{thm:sc_amp_asympB} can be extended to the setting where $\L, n, \log_2\B$ all tend to infinity with the spectral efficiency $S = \L \log_2 \B/n$ held constant (see Remark \ref{rmk:amp_asympB_spectraleff}).
In this asymptotic regime, the result states that for any $S_{\text{AMP}} \leq S < S_{\text{opt}}$, the UER with AMP decoding converges almost surely to 0.
%
\end{remark}

\subsection{Numerical Results}

\begin{figure}[t]
\centering
\includegraphics[width=0.48\textwidth]{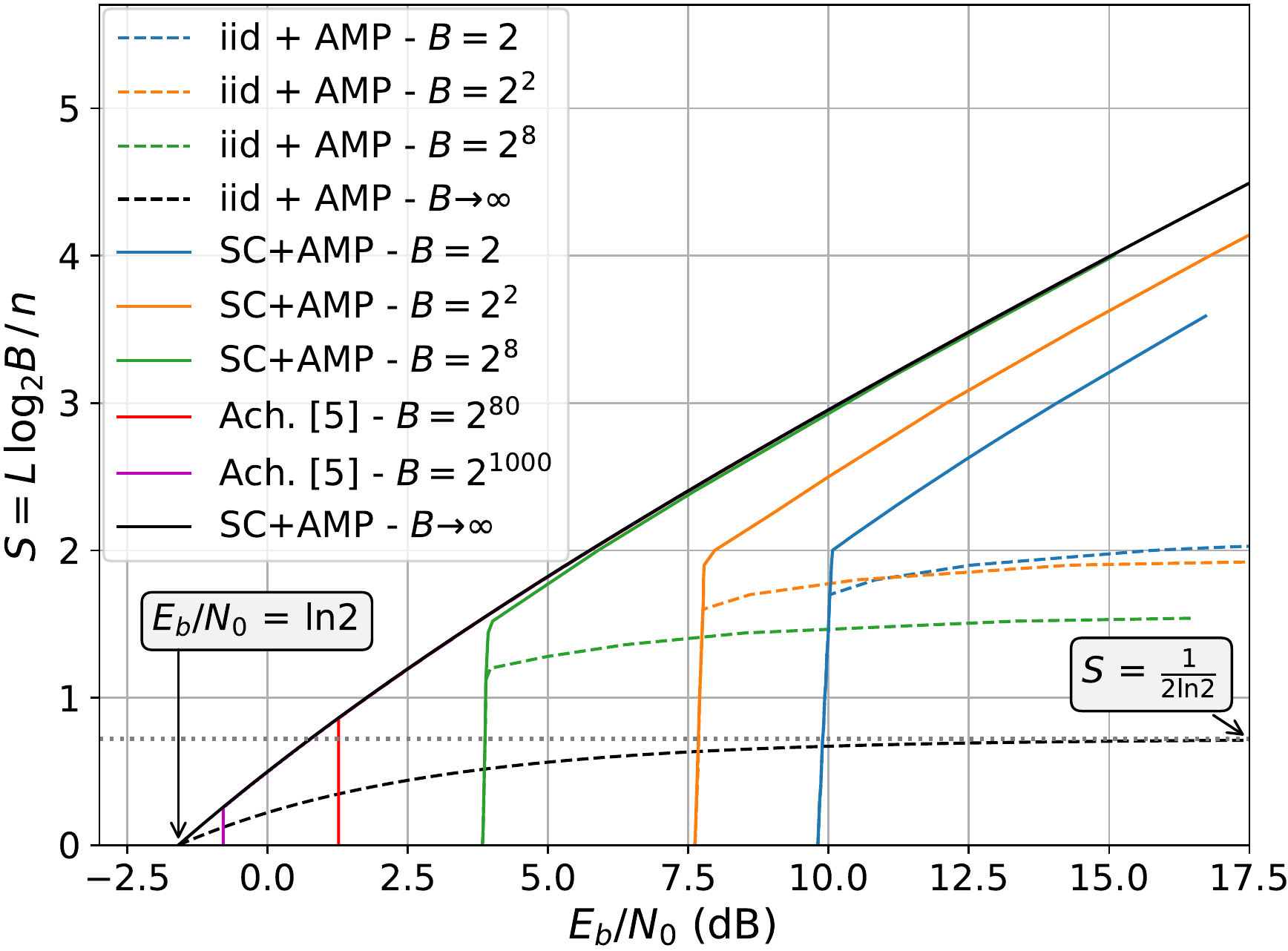}
\caption{Achievable regions of i.i.d. and spatially coupled Gaussian codebooks with AMP decoding at different user payloads ($\log_2 \B$ bits). 
The results at finite $\B$ show the minimum $E_b/N_0$ required to achieve expected UER $\leq 10^{-3}$.
The results for $\B=2^{80}$ and $2^{1000}$ use the achievability bound from \cite {zadik2019isit}.
}
\label{fig:largeM}
\end{figure}

Fig. \ref{fig:largeM} shows the achievable regions of i.i.d. and spatially coupled  Gaussian codebooks with AMP decoding, 
in the large system limit of $\L, n, \log_2 \B$ all tending to infinity with the spectral efficiency $S  = \L\log_2 \B/n$ held constant (Remarks \ref{rmk:amp_asympB_spectraleff} and \ref{rmk:sc_amp_asympB_spectraleff}).
Recall that the user payload is $\log_2 \M= \log_2 \B$ bits.
The dashed black line is the achievable region for i.i.d. Gaussian codebooks and the solid black line for spatially coupled codebooks.
From \eqref{eq:S_BP}, we see that i.i.d. Gaussian codebooks with AMP decoding cannot achieve spectral efficiencies $S \geq \frac{1}{2\ln 2} \approx 0.7213$ (asymptote of the dashed black line).
We note that the solid black line matches the converse bound in \cite{zadik2019isit}
with $\B \to \infty$ and the target expected UER $\epsilon \to 0$.

The solid and dashed black lines split Fig. \ref{fig:largeM} into three distinct regions, which are sometimes referred to as the `easy', `hard', and `impossible' regions of inference in statistical physics \cite{zdeborova2016statistical}:
1) Below dashed black line: achievable with i.i.d. Gaussian codebooks and AMP decoding;
2)  Between solid and dashed black lines: achievable with spatially coupled Gaussian codebooks and AMP decoding (or with i.i.d. Gaussian codebooks and symbol-by-symbol MAP decoding, see Remark \ref{rem:replica});
3) Above solid black line: not achievable by any scheme.

In Fig. \ref{fig:largeM}, we also plot the achievable regions of i.i.d. Gaussian codebooks (dashed lines) and spatially coupled Gaussian codebooks (solid lines) with AMP decoding at several finite payloads $\log_2 \B$ (with $\L, n \to \infty$ and the user density $\mu=L/n$ held constant).
%
%
For $\B=2, 2^{2}, 2^{8}$, we use the same setup as Fig. \ref{fig:gmac_amp_scamp_sim}, and find the smallest $\frac{E_b}{N_0}$ such that the coding scheme achieves UER $\leq10^{-3}$.
For $\B=2^{80}, 2^{1000}$, it is computationally infeasible to evaluate the potential function \eqref{eq:potential_func}, 
so we plot the achievability bound from \cite{zadik2019isit} (red and purple curves).  

For spatially coupled Gaussian codebooks with AMP decoding, the achievable region  gets larger as the user payload increases, but at high spectral efficiencies (e.g. $S>1.5$), the improvement is insignificant after roughly $\log_2 \B =8$ bits. 
Therefore, it is possible to communicate reliably at high spectral efficiencies with near-minimal $\frac{E_b}{N_0}$ even when the user payload is finite.
For i.i.d. Gaussian codebooks with AMP decoding, there is a tradeoff in the achievable region as the user payload increases: 
a lower $E_b/N_0$ is required to communicate reliably at low spectral efficiencies, 
but the maximum achievable spectral efficiency decreases.

\subsection{Implementation at Large User Payloads}\label{sec:large_M_implement}

When i.i.d. or spatially coupled codebooks are used and the user payload $\log_2 \M = \log_2 \B$ bits is large, 
the computational complexity of the AMP decoder is too high for practical use even when DCT based design matrices are used instead of Gaussian ones.
In this section we discuss the computational advantage of using smaller codebooks multiple times and the price paid in terms of a smaller achievable region.
Furthermore, we discuss how introducing modulation to the encoding scheme can reduce the complexity and enlarge the achievable region.

%


A simple idea for reducing the complexity is to encode each user's message using several smaller codebooks instead of a single large codebook, e.g., by using a smaller codebook multiple times. For example, a user payload of $80$ bits can be transmitted using a codebook of size $\B=2^8$ ten times instead of a large codebook of size $\B=2^{80}$ once.
The smaller codebooks can be based on either \iid\ or spatially coupled Gaussian (or DCT) matrices, and the messages can be decoded using an AMP decoder each time.%
%
%
\footnote{ Another possible implementation involves the superposition of codewords from multiple (different) smaller codebooks. For example, user payloads of 80 bits can be transmitted with each user encoding their message with 10 codebooks of size $\B=2^8$ each. The codewords from the 10 codebooks are summed together to form the final user codeword. This method is equivalent to each user encoding their message with a SPARC \cite{joseph2012least}.
This implementation has the same asymptotic achievable region as using smaller codebooks multiple times but has higher decoding complexity. }

The achievable region obtained using this method 
is closely related to the achievable region of the coding scheme where each user encodes a payload of 8 bits with a single codebook of size $\B=2^8$ (the reference coding scheme).
Compared to the reference coding scheme, 
the method described above effectively increases the code length by a factor of 10 due to repeated transmissions,
and the achievable region obtained using this method in terms of the user density $\mu$ versus $E_b/N_0$ trade-off
is the same as that obtained by the reference coding scheme, except the user density $\mu$ is reduced by a factor of ten.%
\footnote{If instead we considered the spectral efficiency $S= (\mu \times \text{user payload})$ versus $E_b/N_0$ trade-off, then the achievable region of the described method would be exactly the same as that of the reference coding scheme. This is because the ten-fold increase in the user payload from 8 bits to 80 bits cancels out the factor of ten reduction in the user density $\mu$.}

\begin{figure}[t]
\includegraphics[width=.48\textwidth]{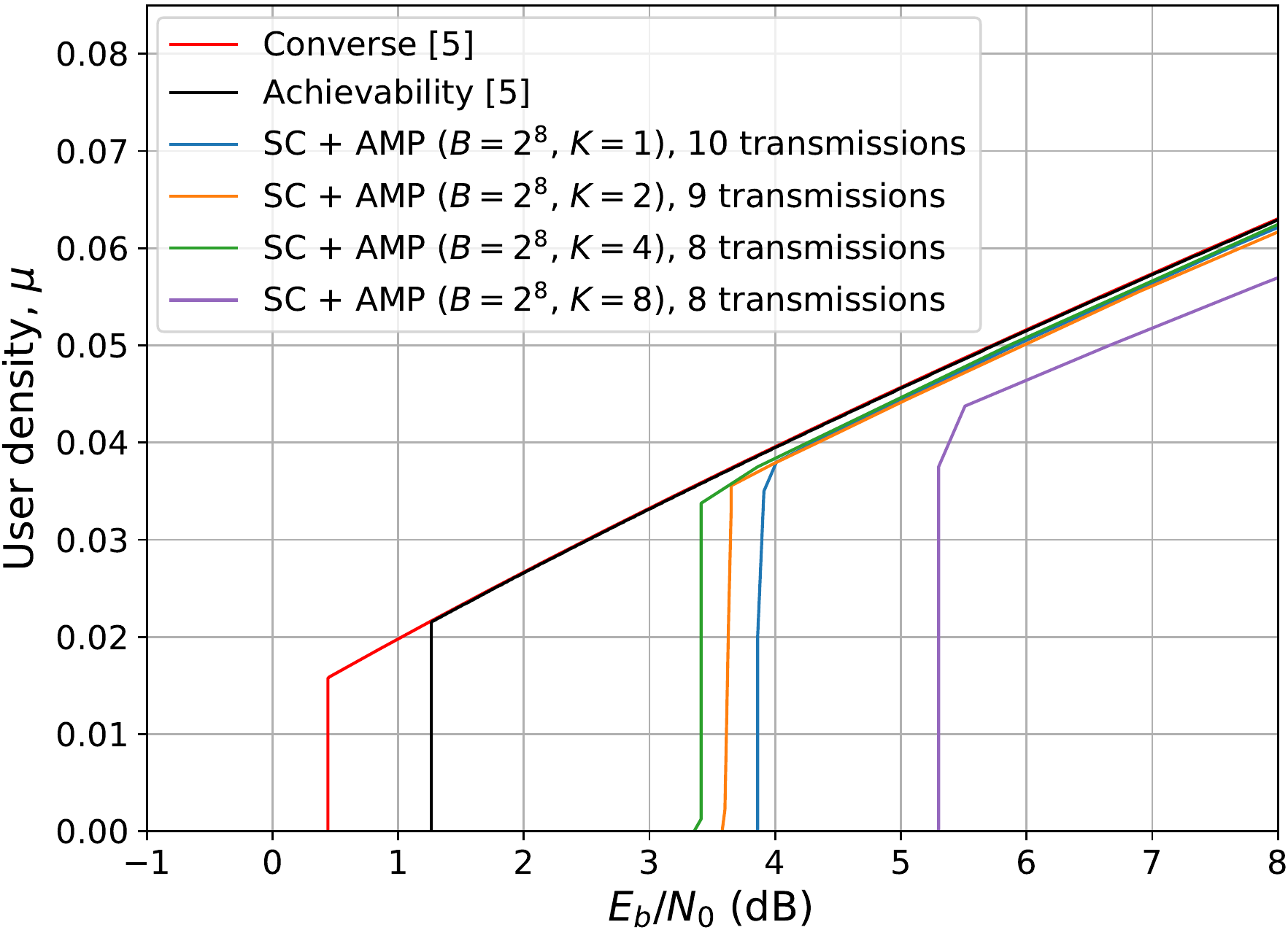}
\caption{Red and black: converse and achievability bounds from \cite{zadik2019isit} for the many-user Gaussian MAC when the per-user payload is 80 bits and the maximum tolerated expected UER is $10^{-3}$.
Blue, orange, green and purple: asymptotic achievable regions of coding methods based on spatially coupled complex Gaussian codebooks (of size $\B=2^8$) with $\K$-ary PSK modulation and AMP decoding for the same maximum tolerated expected UER.
}
\label{fig:k80_complex}
\end{figure}

This asymptotic achievable region of this method is shown in blue in Fig. \ref{fig:k80_complex}, where we assumed that spatially coupled \emph{complex} Gaussian codebooks (of size $\B=2^8$) and AMP decoding are used, and the maximum tolerated expected UER is $10^{-3}$.
The red and black curves plot the converse and achievability bounds from \cite{zadik2019isit} when the user payload is 80 bits and the maximum tolerated expected UER is also $10^{-3}$.
We see that one can still achieve near-optimal $\mu$ versus $E_b/N_0$ trade-offs using these methods at user densities above $\mu \approx 0.04$. 
However, at lower user densities, there is a noticeable gap between the achievability of this method and the achievability of large ($\B=2^{80}$) i.i.d. Gaussian codebooks with ML decoding (black).

We now discuss how to shrink this gap using modulation. 
We consider $\K$-ary phase-shift keying (PSK) modulation in the context of complex Gaussian MACs. 
In the real-valued setting, this modulation technique is restricted to binary PSK.

\textit{Introducing Modulation to Encoding:}
Consider again the example where the user payload is 80 bits. 
If in addition to encoding 8 bits using a size $\B=2^8$ codebook, we encode 1 extra bit in the sign of the chosen codeword (as in Example \ref{ex:distr_binmod}), then the final codeword would encode 9 bits and 80 bits can be sent 
in fewer than 9 transmissions.
In comparison, to send 80 bits without modulation would require 10 transmissions of 8 bits each.
%
%
%
%
For general $\K$-ary PSK modulation, an additional $\log_2 \K$ bits are encoded the phase of the chosen codeword (using Gray coding).
As the modulation factor $\K$ increases, the number of transmissions required (and the complexity) is reduced.
The increase in AMP decoding complexity due to increased modulation is insignificant when $\K\ll \log(\L\B)$ \cite[Sec.~V]{hsieh2021modulated}.
In addition to reducing computational complexity, Fig. \ref{fig:k80_complex} shows that using binary PSK (orange) and 4-PSK modulation (green) increases the achievable region compared to the unmodulated case (blue); 
however, using 8-PSK (purple) significantly decreases the achievable region.

A user's message is decoded in error when either 
the chosen codeword from the user's codebook 
or the phase of the codeword is estimated in error.
We observe empirically that with 8-PSK modulation the phase of the codeword is much more likely to be estimated in error compared to 4-PSK or binary PSK, which results in a smaller achievable region.

\textit{Complex setting:}
To illustrate the effects of PSK modulation,  Fig. \ref{fig:k80_complex} considered a complex Gaussian MAC and coding using complex Gaussian codebooks.
With slight modifications to the AMP decoder, state evolution and potential function, 
Theorems \ref{thm:amp_perf} to \ref{thm:sc_amp_asympB} 
extend directly to the setting where the design matrix $\bA$, message vector $\bx$ and channel noise are complex  (see \cite[Sec.~4.4]{hsieh2021thesis} for more details).
The main takeaway is that complex random linear coding with AMP decoding at user density $\mu$ achieves the same asymptotic UER as real-valued random linear coding with AMP decoding at user density $\frac{\mu}{2}$ when all other parameters are the same, e.g., user payload, $E_b/N_0$ and $p_{\bXsec}$.

\section{Conclusion}

In this paper we considered Gaussian MACs in the asymptotic setting where the number of users $L$ and the code length $n$ both tend to infinity with the user density $\mu = L/n$ held constant, and where the user payload $\log_2 \M$ and energy-per-bit constraint are considered fixed and independent of $n$. We analyzed the asymptotic user error rate (the fraction of user messages decoded in error) achieved by coding schemes based on random linear models (which include i.i.d. Gaussian codebooks and random CDMA) and AMP decoding.
We found that the asymptotic achievable region of a coding scheme based on spatially coupled Gaussian matrices and AMP decoding exceeds that obtained using the achievability bound in \cite{zadik2019isit} and nearly matches the converse bound for a large range of user densities.
The spatially coupled scheme can be interpreted as a block-wise time-division with overlap multiple-access scheme.

We then analysed the performance of these coding schemes as the user payload grows large and found that the interesting asymptotic regime is when $\L,\log_2 \M, n$ all tend to infinity with the spectral efficiency $S=\L\log_2 \M/n$ held constant. 
We also showed that using small random codebooks multiple times to transmit large user payloads can achieve near-optimal trade-offs at large user densities while having lower complexity, and adding modulation (e.g., $\K$-PSK modulation) to the encoding scheme can simultaneously increase the asymptotic achievable region and reduce complexity in such settings.

An interesting open question is how to close the gap between the converse bound and achievable region at low user densities in Fig. \ref{fig:gmac_amp_scamp_sim}. 
Another exciting direction is to explore how ideas such as   block-wise time-division with overlap (via spatially coupled codebooks) and  the generalization of CDMA  (by assigning a codebook of spreading sequences to each user)  can be used to enhance the performance of practical techniques for multiple-access and unsourced random access. For example, in \cite{fengler2021sparcs} the authors suggest spatial coupling as a way to improve the performance of their  unsourced random access scheme based on SPARCs and AMP, and  in \cite{Ebert21CCS}, an outer code is used to induce coupling in the AMP-based random access scheme. Finally, it is also important to study the question of how spatial coupling can be applied in practical  random access settings where the number of users is random and unknown \cite{ngo2021massive}.
%

%

%


%

\appendices
\section{Proof of Lemma \ref{lem:uer_mse}}\label{app:proof_lem_uer_mse}
The existence of the limit in \eqref{eq:uer_mse_bounds} is shown in Theorems \ref{thm:amp_perf} and \ref{thm:threshold_sat}.
For $\ell \in [L]$, we denote by $\boldsymbol{a}_\ell \in \reals^B$  the $\ell$th section of a vector $\boldsymbol{a} \in \reals^{LB}$.
Let $\bx^{t+1}$ be the AMP estimate of $\bx$ after iteration $t+1$ (defined in \eqref{eq:eta_def_p1}).
It was proved in \cite{rush2020capacity} that the MSE of the AMP decoder after iteration  $t\geq0$ converges almost surely to the following limit:
\be\label{eq:amp_mse_conc}
\begin{split}
& \lim_{\L\to\infty }\frac{\|\bx^{t+1} - \bx\|^2}{\L} 
= \frac{1}{\Lc} \sum_{\sfc=1}^{\Lc} \psi^{t+1}_\sfc,
\end{split}
\ee
where the state evolution parameters $\psi_\sfc^t$ for $\sfc\in[\Lc]$ and $t\geq 0$ are defined in \eqref{eq:sc_se_psi}--\eqref{eq:mmse_func}.

To prove \eqref{eq:uer_mse_bounds}, we first notice that for the prior $p_1$, the hard-decision estimator $\hat{\bx}^{t+1}_\ell$ defined in \eqref{eq:amp_final_hard} can equivalently be written as follows. For $j \in \sec(\ell)$:
\begin{equation} \label{eq:hard_dec}
\hat{x}^{t+1}_j=\begin{cases}
 	\ \sqrt{E} \quad &\text{if} \ x^{t+1}_j > x^{t+1}_{i}  \text{ for all }  i \in \sec(\ell)\backslash j,\\
	\ 0 \quad &\text{otherwise}.
\end{cases}
\end{equation}
Here $\bx^{t+1}$ is the AMP estimate computed according to \eqref{eq:sc_amp} and  \eqref{eq:eta_def_p1}. 

Let $j^* \in \sec(\ell)$ denote the index of the unique non-zero entry of $\bx$ in section $\ell\in[\L]$, i.e., $x_{j*} = \sqrt{E}$. From  \eqref{eq:eta_def_p1}, we note that the sum of the entries in each section of  $\bx^{t+1}$  equals $\sqrt{E}$. The decision rule  \eqref{eq:hard_dec} then implies that   $\bx^{t+1}_{j*} \le \sqrt{E}/2$ whenever $\hat{\bx}_\ell^{t+1} \neq \bx_\ell$. Therefore, 
$ \hat{\bx}_\ell^{t+1} \neq \bx_\ell$ implies that
$\|\bx^{t+1}_\ell - \bx_\ell\|^2 \geq E/4$.
Therefore,
\be
\frac{1}{\L} \sum_{\ell=1}^{\L} \mathbbm{1} \{\hat{\bx}_\ell^{t+1} \neq \bx_\ell \}
\leq
\frac{1}{\L} \sum_{\ell=1}^{\L} \frac{4\|\bx^{t+1}_\ell - \bx_\ell\|^2}{E}. \label{eq:indic_sq_error}
\ee
Combining  \eqref{eq:indic_sq_error} with \eqref{eq:amp_mse_conc} yields \eqref{eq:uer_mse_bounds}.


\section*{Acknowledgment}
We thank the reviewers and the Guest Editor for their comments which helped improve the paper.

\ifCLASSOPTIONcaptionsoff
  \newpage
\fi



\bibliographystyle{IEEEtran}
\IEEEtriggeratref{28}
\bibliography{gmac_bib}
\end{document}